\documentclass{ieeeaccess}
\makeatletter
\let\MYcaption\@makecaption
\makeatother
\usepackage{subcaption}
\makeatletter
\let\@makecaption\MYcaption
\makeatother
\usepackage{cite}
\usepackage{amsmath,amssymb,amsfonts}
\usepackage{braket}
\usepackage{siunitx} 
\usepackage{graphicx}
\usepackage{xspace}
\usepackage{qcircuit}
\usepackage{color}
\usepackage{ulem}

\newcommand{\rs}{\textsc{$\sqrt{SWAP}$}\xspace}
\newcommand{\is}{\textsc{$\sqrt{iSWAP}$}\xspace}
\newcommand{\rsx}{\textsc{$\sqrt{SWAP}_{CX}$}\xspace}
\newcommand{\rsv}{\textsc{$\sqrt{SWAP}_{CV}$}\xspace}
\newcommand{\isx}{\textsc{$\sqrt{iSWAP}_{CX}$}\xspace}
\newcommand{\isv}{\textsc{$\sqrt{iSWAP}_{CV}$}\xspace}
\newcommand{\tv}{\textsc{$TOF_{CV}$}\xspace}
\newcommand{\tx}{\textsc{$TOF_{CX}$}\xspace}

\def\BibTeX{{\rm B\kern-.05em{\sc i\kern-.025em b}\kern-.08em
    T\kern-.1667em\lower.7ex\hbox{E}\kern-.125emX}}

\begin{document}
\history{}
\doi{}

\title{Pulse-engineered Controlled-V gate and 
its applications on superconducting quantum device}
\author{
\uppercase{Takahiko Satoh}\authorrefmark{1,2},
\uppercase{Shun Oomura\authorrefmark{1,2}, Michihiko Sugawara\authorrefmark{1,2}, and Naoki Yamamoto}.\authorrefmark{1,2}
}
\address[1]{Quantum Computing Center, Keio University, 
Hiyoshi 3-14-1, Kohoku, Yokohama 223-8522, Japan}
\address[2]{Graduate School of Science and Technology, Keio University, 
Hiyoshi 3-14-1, Kohoku, Yokohama 223-8522, Japan}
\tfootnote{
This work was supported by MEXT Quantum Leap Flagship Program Grant 
Number JPMXS0118067285 and JPMXS0120319794. 
}

\markboth
{Satoh \headeretal: Pulse-engineered Controlled-V gate and 
its applications on superconducting quantum device}
{Satoh \headeretal: Pulse-engineered Controlled-V gate and 
its applications on superconducting quantum device}

\corresp{Corresponding author: Takahiko Satoh (email: satoh@sfc.wide.ad.jp).}

\begin{abstract}
In this paper, we demonstrate that, by employing OpenPulse design kit 
for IBM superconducting quantum devices, the controlled-V gate (CV gate) 
can be implemented in about half the gate time to the controlled-X (CX 
or CNOT gate) and consequently 65.5\% reduced gate time compared to the 
CX-based implementation of CV. 
Then, based on the theory of Cartan decomposition, we characterize the 
set of all two-qubit gates implemented with only two or three CV gates; 
using pulse-engineered CV gates enables us to implement these gates 
with shorter gate time and possibly better gate fidelity than the 
CX-based one, as actually demonstrated in two examples. 
Moreover, we showcase the improvement of linearly-coupled three-qubit 
Toffoli gate, by implementing it with the pulse-engineered CV gate, 
both in gate time and the averaged output-state fidelity. 
These results imply the importance of our CV gate implementation technique, 
which, as an additional option for the basis gate set design, may shorten 
the overall computation time and consequently improve the precision of 
several quantum algorithms executed on a real device. 
\end{abstract}

\titlepgskip=-15pt

\maketitle

\section{Introduction}

There are several type of platforms for implementing quantum computer, 
such as superconducting, ion, and optical devices. 
In this paper, we study the problem of reducing the circuit depth (total 
gate time) in the superconducting quantum device provided by IBM (called 
IBM Quantum), where Qiskit serves as the software development environment. 
Qiskit has two representation languages for designing quantum programs: OpenPulse \cite{qasm} and QASM.

OpenPulse is a language for specifying and physically controlling the pulse level 
of a target quantum gate, that enables introducing a large freedom in circuit 
design. 
As a result, OpenPulse can reduce the execution time through optimal pulse design 
for various type of quantum gates~\cite{gokhale2020optimized,earnest2021pulse}; also it can be applied to generate 
a new gate specific to a particular physical simulation~\cite{stenger2021simulating}. 
Recently, a computational framework has been proposed to aid such synthesis 
problems~\cite{nguyen2021enabling}.

QASM is the language for the circuit design with several quantum gates. 
Physically, each gate is decomposed into a set of precisely calibrated 
gates chosen from the universal quantum gate set \cite{barenco1995elementary, 
brylinski2002universal}. 
The universal gate set used in IBM Quantum is composed of single-qubit 
gates and the Controlled-X (CX, or often called CNOT) gate 
\cite{zulehner2019compiling}. 
The point of taking this fixed gate set is that, because it contains 
only 1 two-qubit interaction gate (i.e., CX gate), the calibration process 
is relatively easy. 
In particular, CX gate can be implemented precisely via the cross 
resonance (CR) Hamiltonian~\cite{kirchhoff2018optimized, cr, riron, sundaresan2020reducing, krantz2019quantum}, with the help of the echo 
scheme and the cancellation pulse technique~\cite{jikken}. 
However, the error rate of CX gate is still much higher than that of 
single-qubit gates~\cite{exp}, due to the longer pulse length (gate time) 
than that of single-qubit gates and the effects of cross-talk 
\cite{murali2020software, mundada2019suppression, sarovar2020detecting}. 
Hence, if a quantum algorithm must be realized on a circuit with 
unnecessarily many CX gates due to the QASM constraint, the accuracy 
of circuit will significantly decrease.

The above-mentioned issue may be resolved by adding some two-qubits gates to the 
default universal gate set composed of single-qubit gates and CX gate. 
In this work, we take the Controlled-V (CV) gate whose matrix 
representation in the computational basis is given by 
\begin{equation}
    CV =\left[\begin{array}{rrrr}
1&0&0&0\\
0&1&0&0\\
0&0&\frac{1+i}{2}&\frac{1-i}{2}\\
0&0&\frac{1-i}{2}&\frac{1+i}{2}
\end{array}\right], 
\end{equation}
which readily leads to the relation $CV^2=CX$. 
Note that, in the CX-based default implementation of CV gate on QASM, one needs 
2 CX gates to create the two-qubits interaction process as shown in 
Fig.~\ref{fig:circuit_cv}.

\Figure[htb][width=240pt]{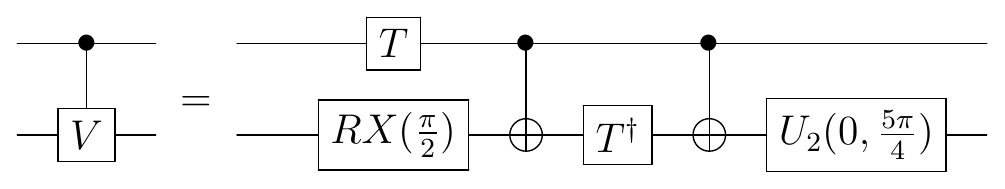}
{\label{fig:circuit_cv}Circuit diagram of CV gate in QASM-based implementation.}

The main reason for choosing CV gate is its potential ability to reduce 
the gate time in several QASM-based quantum algorithms. 
The point is that, by using OpenPulse, we can effectively 
implement CV gate by just halving the pulse length of the CR pulse used 
for generating CX gate, as suggested by the relation $CV^2=CX$. 
That is, the gate time of pulse-engineered CV gate is half that of CX 
gate, while the QASM-based CV gate shown in Fig.~\ref{fig:circuit_cv} 
needs the gate time at least twice that of CX.
Therefore, if some CX gates on a quantum circuit can be replaced with 
the same or less number of CV gates, the total gate time of this circuit 
is reduced and thereby the accuracy of the circuit will be improved. 
A typical example is the Toffoli gate; it needs at least 6 CX gates to 
implement if CX is only given to us, but it can be implemented using 
2 CX and 3 CV gates if CV gate is further available 
\cite{barenco1995elementary, divincenzo1998quantum, cs}.

This paper is organized as follows. 
In Section II, we describe how to implement CV gate using OpenPulse and 
then show the experimental result; the gate time of the pulse-engineered 
CV gate is shortened by 65.5\% and the gate fidelity is improved by 0.66\%, 
compared to the default QASM-based implementation of CV gate. 
In Section III, we first use the theory of Cartan decomposition to 
characterize the set of all two-qubit gates implemented with only two or 
three CV gates; because the pulse-engineered CV gate can be implemented 
with shorter gate time, those two-qubit gates can also be implemented 
with shorter gate time and possibly better gate fidelity than the 
default CX-based one. 
Actually, we show the experimental demonstration to generate \(\sqrt{iSWAP}\) 
and \(\sqrt{SWAP}\) using the pulse-engineered CV gates and confirm that, 
in both cases, the gate fidelity is improved thanks to the shorter gate time. 
In Section IV, we showcase an efficient method for implementing a  
linearly-coupled three-qubit Toffoli gate using the pulse-engineered CV 
gate.

\section{Pulse-engineered CV gate}
\label{sec_cv}
\subsection{Cross resonance interaction}
On IBM Quantum devices, the cross resonance (CR) interaction is used to couple 
two qubits \cite{cr}, by irradiating the control qubit with a microwave pulse 
at the transition frequency of the target qubit. 
The microwave pulse has a Gaussian-square type envelope in the default setup; 
see Appendix~A. 
Under some approximation, we obtain the following model CR Hamiltonian 
\cite{jikken,cs, riron, sundaresan2020reducing}: 
\begin{align}
    H_{CR} =& \sum_{P = I, X, Y, Z}\frac{\omega_{ZP}(A,\phi)}{2}Z\otimes P \notag\\
    &+ \sum_{Q = X, Y, Z}\frac{\omega_{IQ}(A,\phi)}{2}I\otimes Q,
    \label{hamii}
\end{align}
where the qubit ordering is control\(\otimes\)target. 
\(\omega_{ZP}\) and \(\omega_{IQ}\) represent the interaction strength, which 
are functions of the amplitude $A$ and the phase \(\phi\) of the microwave 
pulse. 
Note that the CR Hamiltonian is valid under the condition that  
the microwave pulse with transition frequency of the target qubit is
irradiated to the control qubit. 
In the absence of noise, the two qubits are driven by the unitary operator 
\begin{align}
    U_{CR}
    = \exp(-itH_{CR}). 
    \label{unii}
\end{align}

\subsection{Pulse-engineered CX and CV gates}
Let us define the general two-qubit unitary operator 
\begin{equation}
\label{eq_2q_ope}
    [DE]^{\theta}=\exp \Big( -i\pi\frac{\theta}{2}D\otimes E \Big), 
\end{equation}
where $D$ and $E$ are arbitrary single-qubit operators. 
With this notation, the CX gate is represented as \cite{cr}: 
\begin{equation}
\label{cnotdeco}
    CX = [ZI]^{1/2}[ZX]^{-1/2}[IX]^{1/2}. 
\end{equation}
That is, the two-qubit operation required to form the CX gate can only be 
served by the $Z\otimes X$ Hamiltonian. 
However, the CR Hamiltonian (\ref{hamii}) contains terms other than 
$Z\otimes X$ term, which thus should be eliminated by some means for 
implementing the CX gate via the CR Hamiltonian. 
This goal can be achieved, by using the echo sequence pulse scheme and 
applying a direct cancellation pulse on the target qubit as illustrated 
in Fig.~\ref{cnot}; 
in other words, these techniques are effectively used to generate the 
unitary evolution driven by the effective Hamiltonian, ${\tilde H}_{ZX}$, 
composed of only the $Z\otimes X$ term \cite{pro, sundaresan2020reducing}. 
In general, one can implement the unitary operator $[ZX]^{\theta}$ driven 
by the effective Hamiltonian ${\tilde H}_{ZX}$, by setting the interaction 
strength in terms of the pulse duration $t$ as 
$\theta = \omega_{ZX}(A,\phi) t/\pi$; 
\begin{equation}
\label{unitary}
\begin{split}
    [ZX]^{\theta} 
    &= {\tilde U}_{ZX} = \exp(-i \pi t {\tilde H}_{ZX}), \\
    {\tilde H}_{ZX}
    &= \frac{\omega_{ZX}(A,\phi)}{2}Z\otimes X. 
 \end{split}
\end{equation}
For the CX gate case, the two-qubit interaction time $t_{CX}$ should be 
$t_{CX} = \pi/2 \, \omega_{ZX}(A,\phi)$ to realize $\theta = -1/2$. 

\Figure[htb][width=220pt]{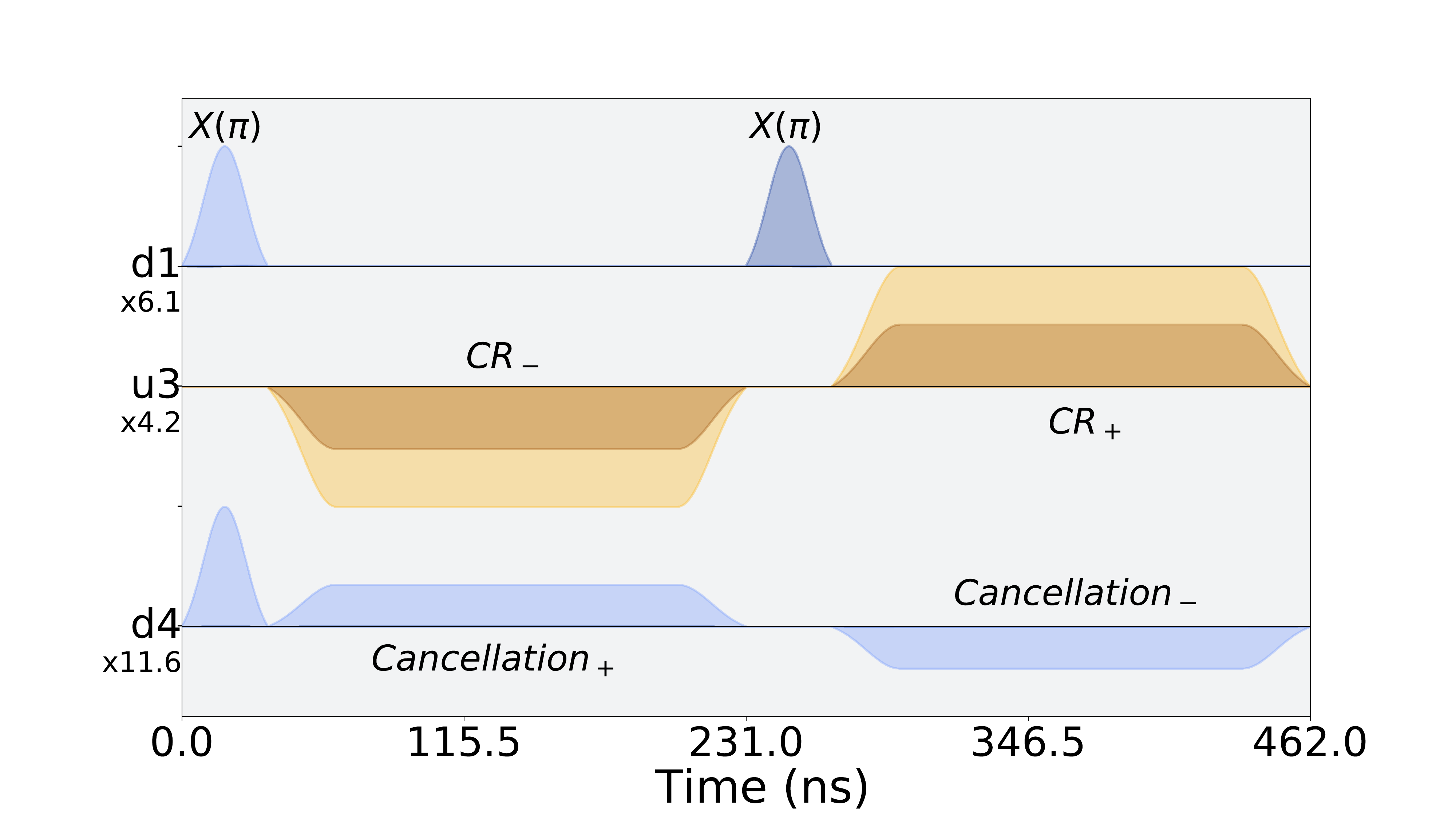}
{Pulse schedule of CX gate for the control qubit 1 and the target qubit 4,  
implemented on IBM Q Toronto. 
Here, d1 and d4 denote the drive channels for local operations of qubits 
1 and 4, while u3 is the control channel for CR-pulse responsible for 
the two-qubit interactions. 
$CR_+$ and $CR_-$ are the CR pulse shapes implementing $[ZX]^{1/2}$ on u3;
d4 is the channel serving as the cancellation pulse. 
Two $\pi$-pulses on d1 placed before and after the first CR-pulse are used to 
realize the echo-scheme. 
The first Gaussian pulse on d4 corresponds to $[IX]$ in Eq.~\eqref{cnotdeco}, 
whereas $[ZI]$ is implemented without actual pulse-irradiation.
\label{cnot}
}

Next, from Eq. (\ref{cnotdeco}) and the fact that $IX$, $ZX$, and $ZI$ 
commute with each other, one can see that CV gate is decomposed as
\begin{equation}
    CV = [ZI]^{1/4}[ZX]^{-1/4}[IX]^{1/4}.
\end{equation}
In the present work, we directly implement \([ZX]^{-1/4}\) part using 
OpenPulse, without decomposing this gate into multiple CX gates.
As expected from Eq. (\ref{unitary}), the interaction strength \(\theta\) 
of the two-qubits interaction part \([ZX]^{\theta}\) is proportional to 
the duration of CR pulse, as far as the effective Hamiltonian stands. 
Thus, we can create \([ZX]^{-1/4}\) by taking the duration of the CR pulse 
\(t_{CV}\) as
\begin{align}
 t_{CV} = \frac{\pi}{4\omega_{ZX}(A,\phi)}, 
\end{align} 
which is half the value of calibrated CX gate's CR pulse duration. 
The CR pulse envelope is a GaussianSquare pulse, i.e. a square pulse with Gaussian-shaped rising and falling edges~\cite{pro} (see also Appendix~\ref{sec_envelope}).

Note that, in all experimental demonstration shown in this paper, we keep the 
basic structure of the pulse schedule and amplitude parameters, for the combined 
CR and cancellation pulses in Fig.~\ref{cnot} unchanged, whereas we replace the local 
gate parameters for $[IX]^{1/2}$ and $[ZI]^{1/2}$ in the CX pulse definition 
with those of $[IX]^{1/4}$ and $[ZI]^{1/4}$; moreover, the CR pulse duration 
is changed to the value corresponding to $[ZX]^{1/4}$.

\subsection{Experimental environment}
In the present work, we used the 0th, 1st, and 4th qubits of ibmq\_toronto, 
as shown in Fig.~\ref{processor}. 
Single-qubit gate operations on qubits 0, 1, and 4 are realized by the 
microwave irradiation to the drive-channel, d0, d1, and d4, respectively, 
whereas the CR-pulses for the two-qubit interactions between qubits 0 and 1, 
and qubits 1 and 4 are applied to the control channels, u0 and u3. 
Each experiment demonstrated in this paper was conducted 8192 times 
(meaning that 8192 measurement was performed for each circuit). 
There exist measurement errors that accidentally flips the detected bit; 
we applied the readout error mitigation technique~\cite{Qiskit} to fix 
this error. 
We list the single-qubit gate error and the readout error of the device in 
Table~\ref{oneperformance}. 
Also the two-qubit CX gate errors are 1.065\% and 1.5969\% for the 
0-1 qubits pair and 1-4 qubits pair, respectively. 

\Figure[htb][width=4.5cm]{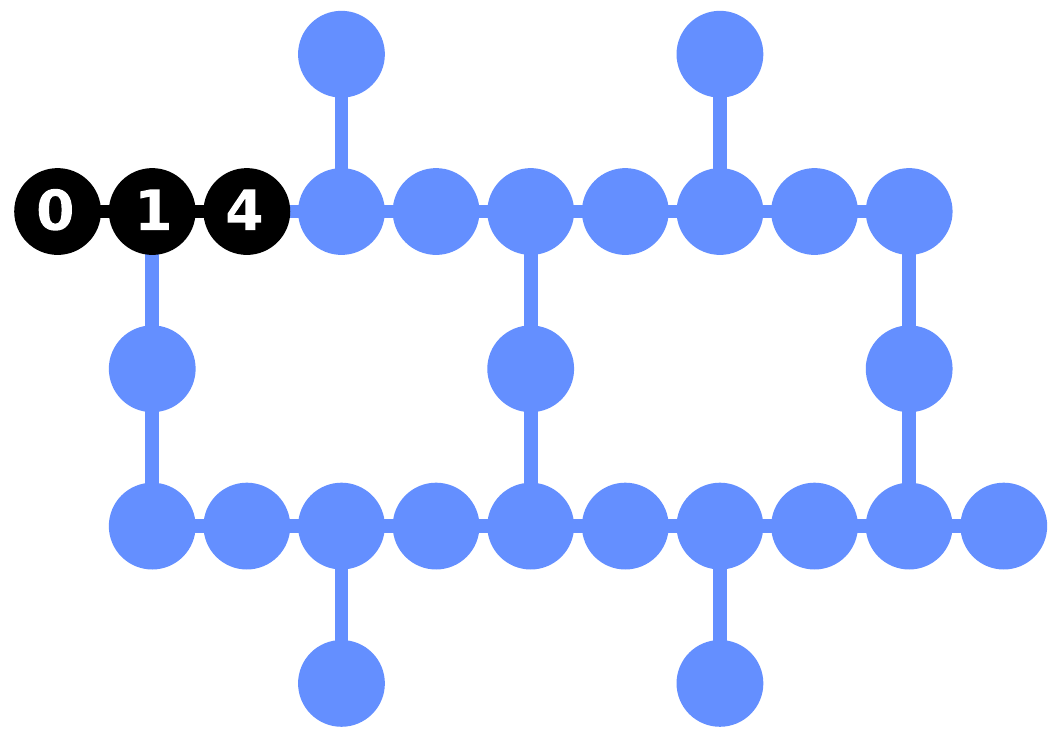}
{The coupling map of ibmq\_toronto processor. 
We used qubits 0, 1, and 4 for the experiments.\label{processor}}

\begin{table}[h]
 \centering
 \caption{Qubit performance of ibmq\_toronto processor.}
  \begin{tabular}{ccc} 
    qubit&$\sqrt{X}$ gate error & Readout error \\\hline
    0&2.78e-4&5.69e-2\\ 
    1&6.97e-4&4.05e-2\\ 
    4&3.42e-4&7.30e-2
  \end{tabular}
 \label{oneperformance}
\end{table}

\subsection{Experimental Results}
We implemented the gate \eqref{unitary} with several values of the pulse 
duration \(\tau_d\) of the two CR pulses, which correspond to $CR_-$ 
and $CR_+$ shown in Fig.~\ref{cnot}, from 45.5 ns to 161 ns. 
For each \(\tau_d\) we test the following trial CV gate: 
\begin{equation}
\label{trial CV}
    CV_{\rm trial}(\tau_d) = [ZI]^{1/4}[ZX]^{\theta(\tau_d)}[IX]^{1/4},
\end{equation}
where $\theta(\tau_d) = -\tau_d/4 t_{CV}$. 
Note that the duration for realizing the CX gate is 196 ns (see Appendix~\ref{sec_cxpulse} for details); hence, from the relation $CV^2=CX$, 
ideally $\tau_d$ would be identical to $\tau_{CV}=98 = 196/2$ ns to realize CV gate.
We make this duration adjustment only for the flat-top part, and the 
Gaussian flanks are fixed.
We applied the quantum process tomography (QPT) to construct the trial 
CV gate, to evaluate its gate fidelity $F_p$ to the ideal CV gate 
\cite{choi, schumacher1996sending, havel2003robust}. 
Note that we can use interleaved randomized benchmarking~\cite{cs} or {\it randomized\_benchmarking} function in the Qiskit libraries~\cite{Qiskit}, to estimate the gate fidelity.

\Figure[ht][width = 230pt]{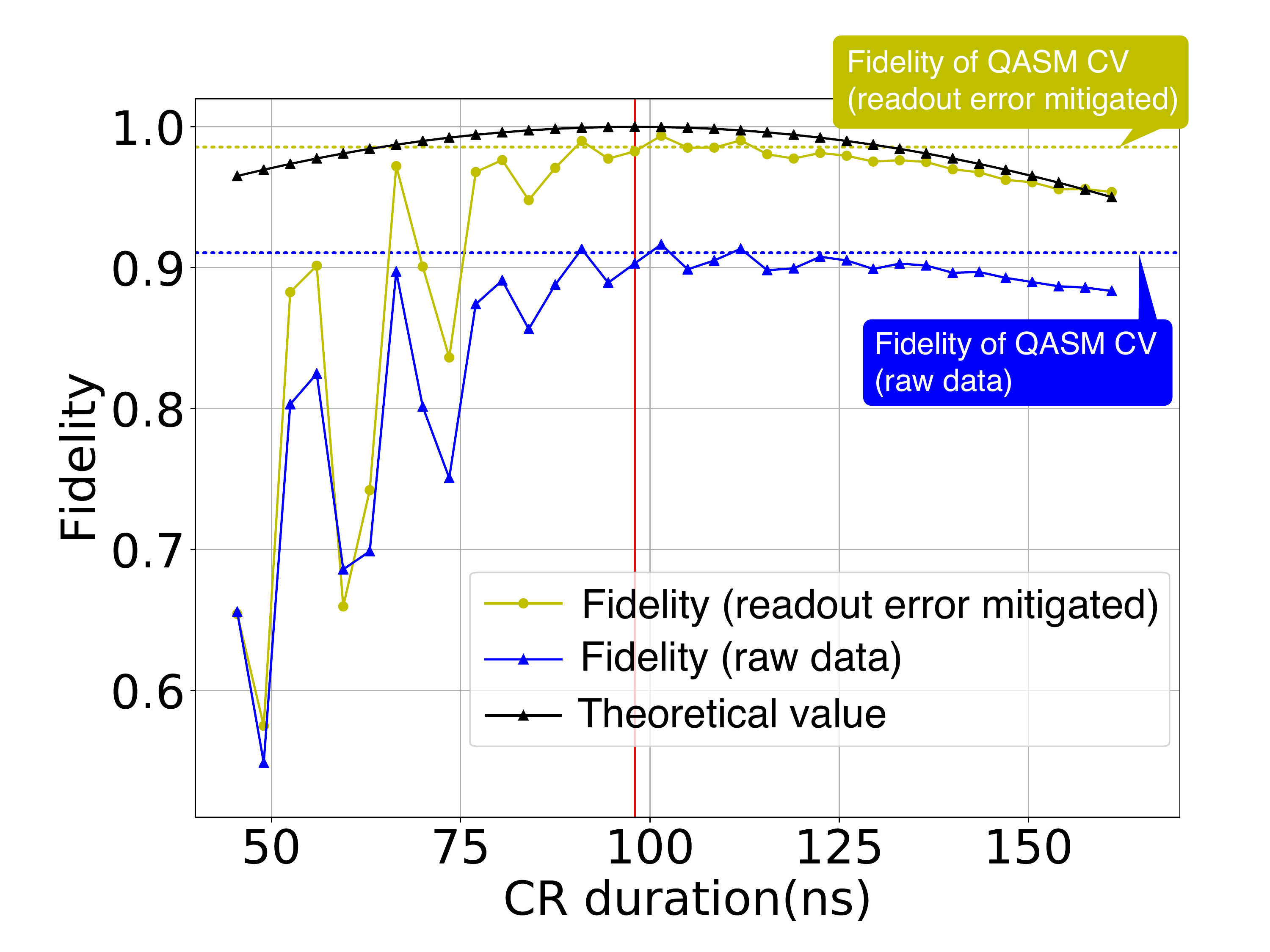}
{Gate fidelity of the trial CV gate \eqref{trial CV} to the ideal CV gate, 
as a function of the duration of CR pulse. 
Red vertical line denotes half duration of CR pulse in CX pulse schedule.
The black line represents the theoretically calculated gate fidelity between 
the exact CV gate and the trial CV gate (9). 
Here, the physical control and target qubit is the 0-th and the 1st one depicted 
in Fig.~\ref{processor}, respectively. 
\label{cv_fidelity}}

Figure~\ref{cv_fidelity} shows the gate fidelity of the trial CV gate 
\eqref{trial CV} as a function of the CR pulse duration $\tau_d$, with 
and without the readout mitigation; 
these are the averages of three experimental results conducted three 
different days. 
The black line represents the theoretically calculated gate fidelity between 
the exact CV gate and the trial CV gate~\eqref{trial CV}, as a function of the CR duration; 
in the latter, $[ZX]^\theta$ can be analytically calculated using Eq.~\eqref{eq_2q_ope}, and $\theta(\tau_d)$ linearly increases with respect to $\tau_d$. 
Also for reference, the gate fidelity of the CV gate implemented in the 
QASM format (denoted as QASM CV) are shown. 
First, note that the readout error-mitigation works well and gives better 
fidelity values compared to the raw (unmitigated) results. 
The mitigated fidelity of CV gate implemented with OpenPulse (denoted 
as Pulse CV) takes the maximum value 99.23\% (averaged value for three 
different days) at the CR duration $\tau_d=101.5$ ns, which is close to the 
expected value $\tau_{CV} = 98$ ns, i.e., half the duration of CR pulse of 
the calibrated CX gate. 
Throughout all three different experiments, the maximum value is taken at 
101.5 ns, which indicates that the optimal pulse duration is robust against 
calibration change. 
Another important finding is that the maximum value 99.23\% is 0.66\% higher 
than that of the CV gate fidelity achieved via the default QASM-based 
implementation using 2 CX gates. 

\begin{figure}[ht]
    \begin{minipage}{\hsize}
    \centering
    \includegraphics[clip,width = 225pt]{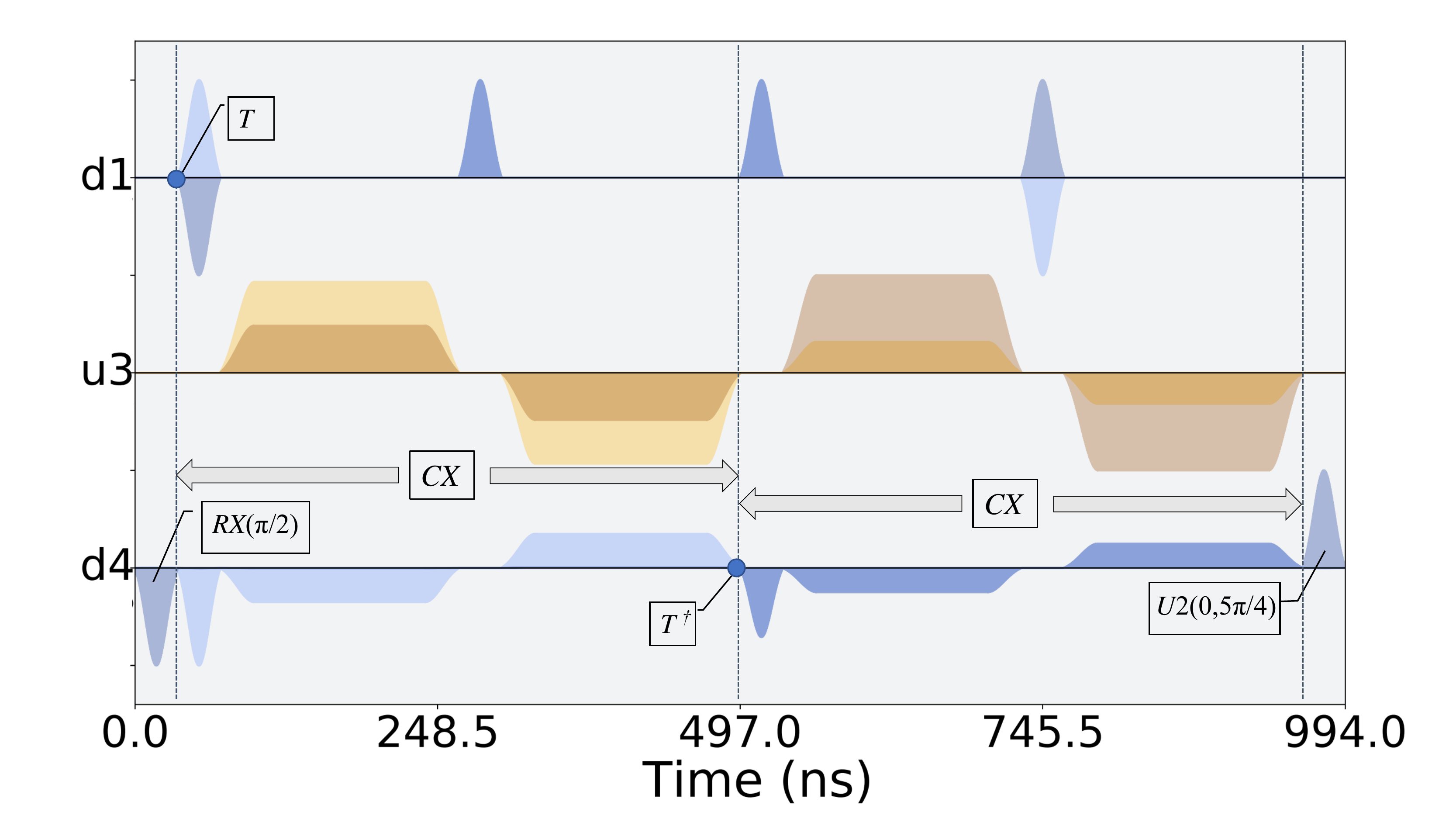}
    \end{minipage}
    \centerline{(a)}
    \begin{minipage}{\hsize}
    \includegraphics[clip,width = 240pt]{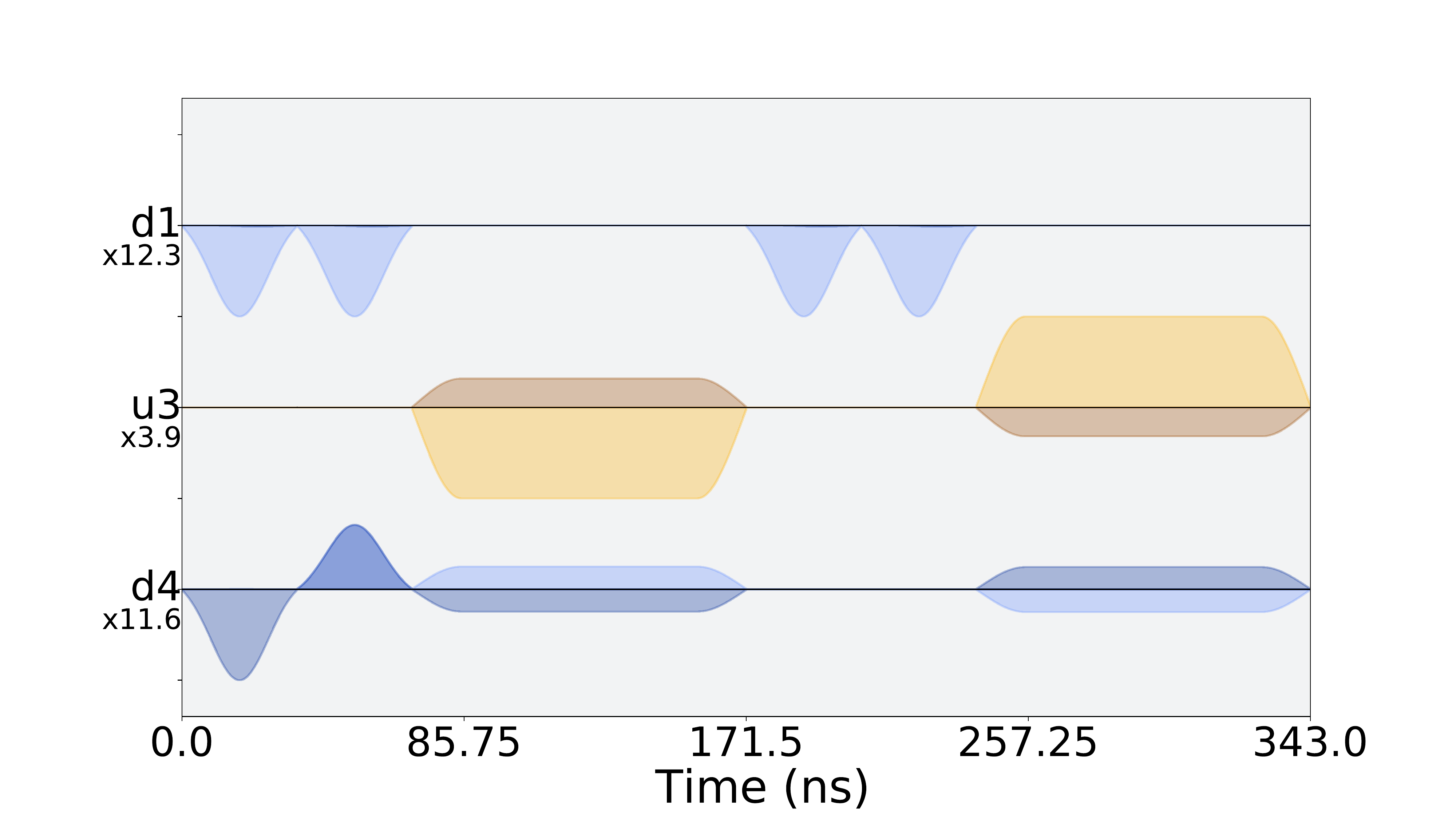}
    \end{minipage}
    \centerline{(b)}
    \caption{(a) Pulse sequence for CV gate in the QASM implementation. 
    (b) Pulse sequence for CV gate implemented by OpenPulse with the CR pulse 
    duration 101.5 ns. }
    \label{cv_pulse}
\end{figure}

Figure~\ref{cv_pulse} shows the actual pulse sequence of CV gate implemented 
in (a) the default QASM format with 2 CX gates (see Fig.~1) and (b) OpenPulse 
with the optimal pulse duration 101.5 ns. 
The total gate time of CV gate is 994 ns for the former, while it is 343 ns 
for the latter. 
Hence the present OpenPulse-based implementation achieves 65.5\% 
reduction in the total gate time of CV gate, compared to the default one 
(a), in addition to 0.66\% improvement in the gate fidelity. 

\section{Two-qubit gate design with CV Gates}
\label{sec_twoqgate}
Arbitrary two-qubit gates can be implemented with three CX gates~\cite{geo, vidal2004universal}.
However, generating two-qubit interactions only with CX gates can unnecessarily 
prolong the gate time. 
In this section, we study the set of two-qubit gates that can be configured 
with up to three CV gates instead of the same number of CX gates, based on 
the theory of Cartan decomposition. 
In particular, we consider \rs gate and \is gate as examples; they can be 
implemented with three and two CV gates, respectively, and thus 
the resulting gate-time is obviously shortened compared to the default 
CX-based implementations. 
We have also experimentally confirmed that the gate fidelity of those 
CV-based gates is superior to that of the CX-based one. 

\subsection{Cartan decomposition}
The Cartan decomposition proves that an arbitrary two-qubit unitary operation 
\(U\in SU(4)\) can be represented in the form
\begin{equation}
\label{cartan}
    U = k_1\exp\{\frac{i}{2}(aX\otimes X+bY\otimes Y+cZ\otimes Z)\}k_2, 
\end{equation}
where \(k_1,k_2\in SU(2)\otimes SU(2)\) are local single-qubit operations. 
When two-qubit unitaries $U$ and $V$ are connected through \(U = k_1 V k_2 \), 
we call that \(U\) and \(V\) are locally equivalent.

\begin{figure}[ht]
\centering
\includegraphics[clip,width = 7.2cm]{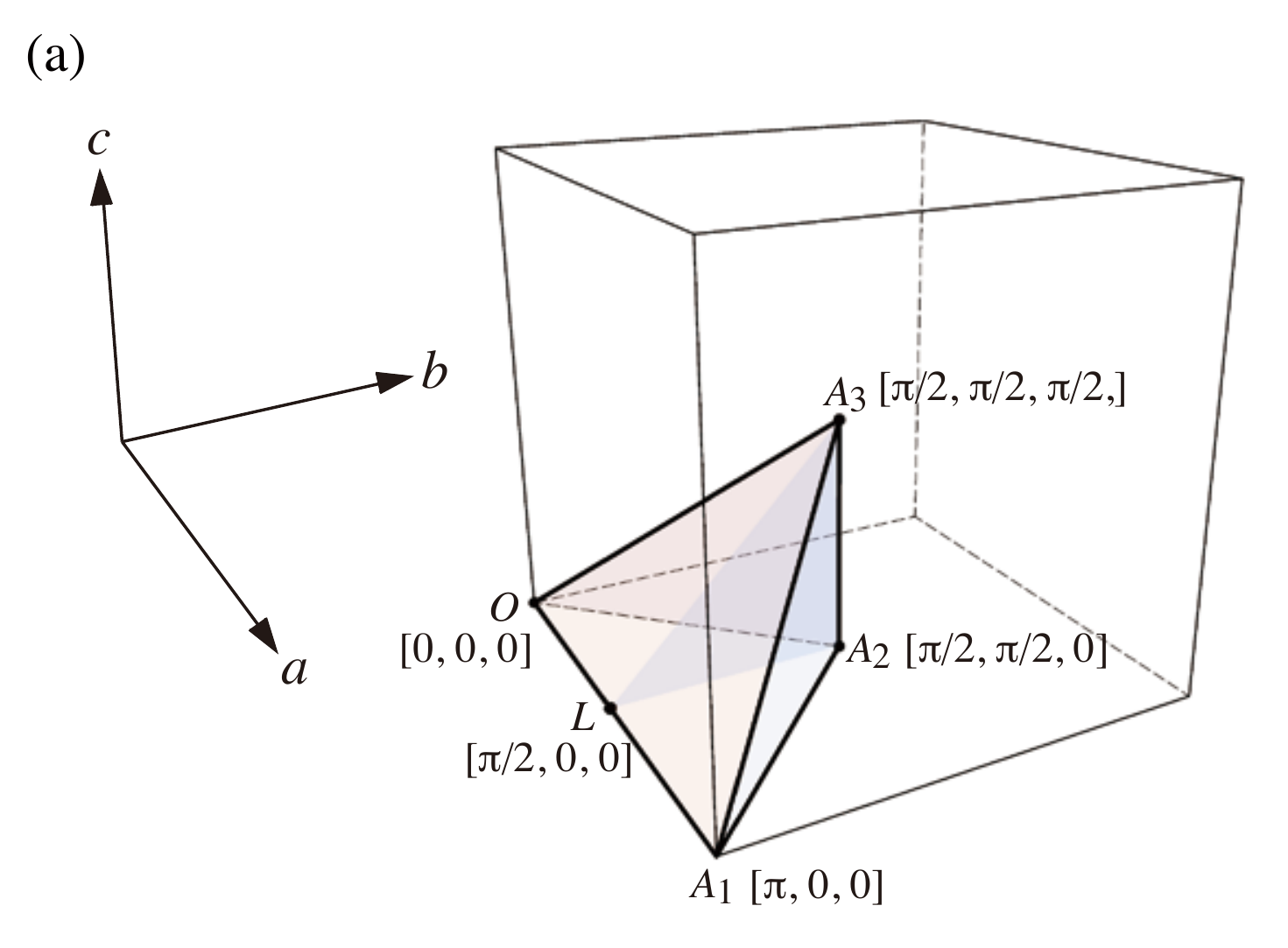}
\includegraphics[clip,width = 4.0cm]{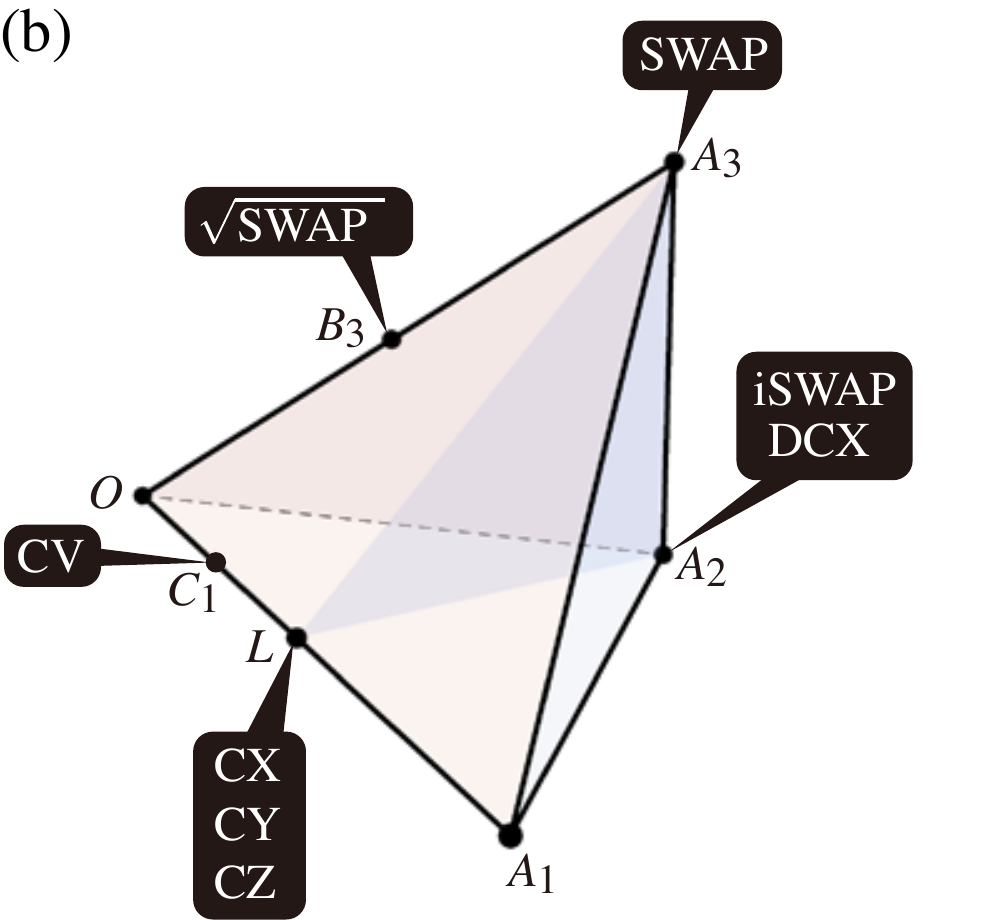}
\includegraphics[clip,width = 4.3cm]{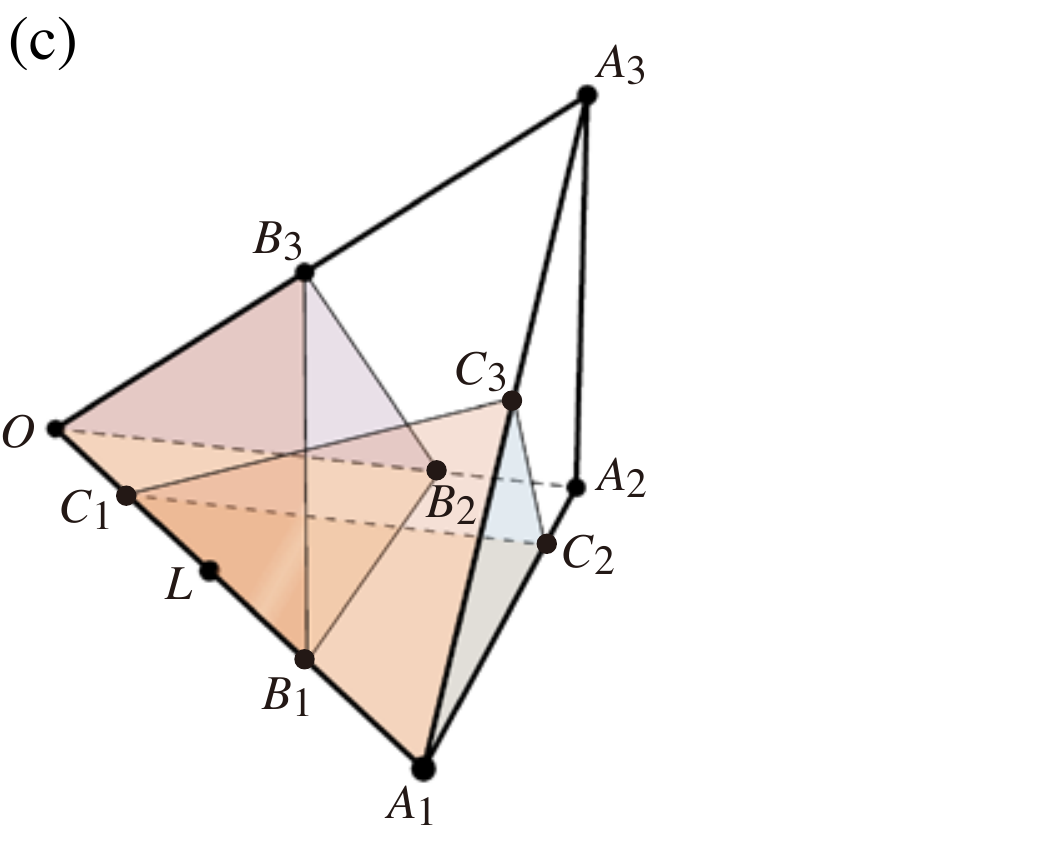}
\caption{(a) Weyl chamber (tetrahedron \(OA_1A_2A_3\)) contains all the 
locally equivalent class of two-qubit operations, with the exception of 
points on its base (see the caption of Fig. \ref{2D}). 
(b) Five important points in the Weyl chamber; at each point typical locally 
equivalent gates are indicated. 
(c) Colored area, i.e., the union of tetrahedra \(OB_1B_2B_3\) and 
\(A_1C_1C_2C_3\) in the Weyl chamber, shows the set of two-qubit gate 
realized with three CV gates. 
All the points in the figures are defined as 
\(B_1 =[3\pi/4,0,0]\), \(B_2 =[3\pi/8,3\pi/8,0]\), \(B_3 =[\pi/4,\pi/4,\pi/4]\), \(C_1 =[\pi/4,0,0]\), \(C_2 =[5\pi/8,3\pi/8,0]\), \(C_3 =[3\pi/4,\pi/4,\pi/4]\).
}
\label{Weyl-Chamber}
\end{figure}

The Cartan decomposition is directly used to construct Weyl chamber that 
provides a clear view of geometric structure of the set of all non-local 
two-qubit gates. 
The Weyl chamber is illustrated as the tetrahedron \(OA_1A_2A_3\) in 
Fig.~\ref{Weyl-Chamber}(a); 
the point $[a, b, c]$ represents a locally equivalent class of two-qubit 
gate \cite{geo, opt}. 
Shown in Fig.~\ref{Weyl-Chamber}(b) are particularly important points 
corresponding to familiar two-qubit gates, 
$L=[\pi/2,0,0]$ for \{CX,~CY,~CZ\}, 
$A_2=[\pi/2,\pi/2,0]$ for \{DCX,~iSWAP\}, 
$A_3=[\pi/2,\pi/2,\pi/2]$ for SWAP, and 
$B_3=[\pi/4,\pi/4,\pi/4]$ for \rs. 
Note from Eq.~\eqref{cnotdeco} that CX is locally equivalent to 
$[ZX]^{-1/2}$, which is further locally equivalent to $[XX]^{-1/2}$ 
and thus identified by $L=[\pi/2, 0, 0]$. 
From this view, it is clear that CV corresponds to $C_1=[\pi/4, 0, 0]$.

\begin{figure}[ht]
    \begin{minipage}{\hsize}
    \centering
    \includegraphics[clip,width = 7cm]{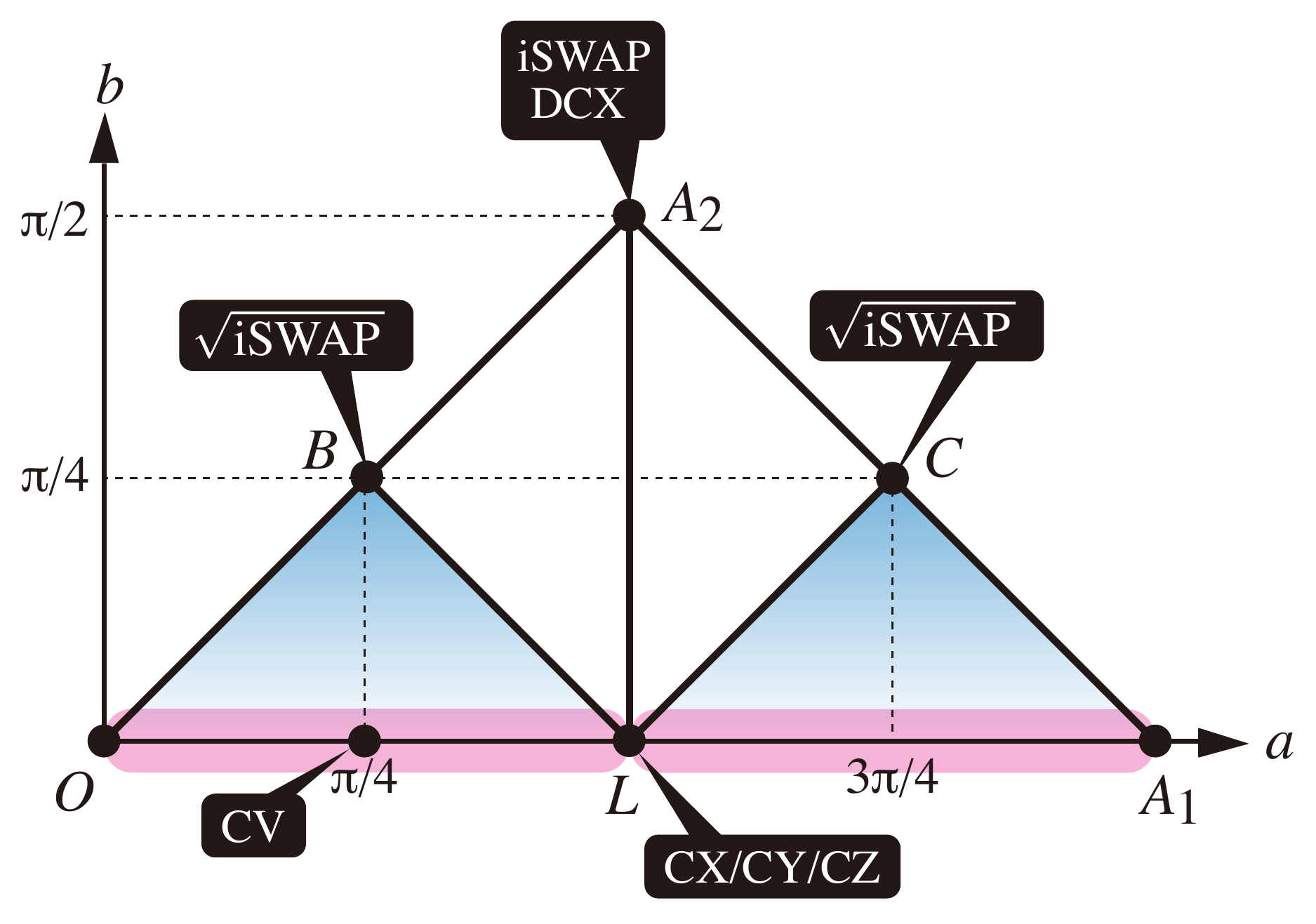}
    \end{minipage}
\caption{Area of two-qubit gates generated with 2 CX or CV gates. 
2 CX gates can generate any gate represented by the point in the 
triangle area \(OA_1A_2\), which is the base of Weyl chamber. 
The blue region represents the set of gates that 2 CV gates can 
generate. 
The triangle areas \(OLB\) and \(A_1LC\) are locally equivalent; 
in particular, \(B =[\pi/4,\pi/4,0]\) and \(C =[3\pi/4,\pi/4,0]\) 
are equivalent points corresponding to $\sqrt{\mbox{iSWAP}}$ gate. 
The lines $OL$ and $LA_1$ (overlaid with pink belt) correspond to 
the set of controlled-$U$ gates. 
}
\label{2D}
\end{figure}

A particularly useful result provided by this geometric picture is 
that \(n ~ (\geq 3)\) times repetition of \([\gamma,0,0]\) with 
$\gamma\in (0, \pi/2]$ can create an arbitrary two-qubit gate 
$[a,b,c]$ that satisfies the following condition: 
\begin{equation}
\label{CU3D}
0\leq a+b+c\leq n\gamma, ~ a-b-c\geq \pi-n\gamma. 
\end{equation}
This equation implies that $n=3$ operations of CX (or any of locally 
equivalent gate to \([\pi/2, 0, 0]\)) with appropriate local gates can span 
the entire area of Weyl chamber, i.e., tetrahedron \(OA_1A_2A_3\); 
that is, as is well known, 3 CX gates can generate arbitrary two-qubit 
unitary gates. 
Similarly, by using two \([\gamma,0,0]\) gates, we can create arbitrary 
two-qubit gate $[a,b,0]$ that satisfies the following 
condition:
\begin{equation}
\label{CU2D}
0\leq a+b\leq 2\gamma, ~ a-b\geq \pi-2\gamma.
\end{equation}
Thus, two CX gates can generate any two-qubit gate represented by the 
point inside the triangle $OA_1A_2$, which corresponds to the base of 
the Weyl chamber (see Fig.~\ref{2D}). 

\subsection{Configurable CV-based two-qubit gates}
We can now characterize the set of two-qubit gates generated by two or 
three operations of CV gate represented by $C_1=[\pi/4, 0, 0]$. 

First, Eq.~\eqref{CU2D} with $\gamma=\pi/4$ indicates that 2 CV gates 
can generate any unitary gate represented by the point in the locally 
equivalent areas \(OLB\) and \(A_1LC\) illustrated in Fig.~\ref{2D} 
\cite{opt}. 
These areas are included in the triangle \(OA_1A_2\). 
Hence, there exist gates such that 2 CX gates can generate while 2 CV 
gates cannot, such as DCX 
(Double-CX gate, i.e., a 2-qubit gate composed of 
two back-to-back CX gates with alternate controls) or equivalently iSWAP represented by \(A_2=[\pi/2, \pi/2,0]\). 
However, there are still many useful two-qubit gate in \(OLB\) and 
\(A_1LC\), and it is thus important to have the pulse-engineered CV gate 
for generating those gates with significantly shorter time and possibly 
better gate fidelity than the case using the default QASM-based 
implementation with only CX. 
For example, the controlled-$U$ gate plays an essential role in 
several quantum algorithms such as 
Quantum Fourier Transform; fortunately, an arbitrary 
controlled-$U$ gate is specified by the point \([\gamma,0, 0]\) on the 
line $OL$ or $A_1 L$ and thus can be generated using two CV gates. 

Second, Eq.~\eqref{CU3D} with $n=3$ and $\gamma=\pi/4$ elucidates the 
set of two-qubit gates that can be generated with 3 CV gates, which 
is depicted in the colored area in Fig.~\ref{Weyl-Chamber}(c). 
We can expect the same advantage as the 2 CV case, in implementing 
some two-qubit gates contained in this area via three pulse-engineered 
CV gates. 

\subsection{Efficient implementation of \is and \rs via pulse-engineered 
CV gates}
Here we show an experimental demonstration to implement the following 
2 two-qubit gates via the pulse-engineered CV gates. 
That is, we consider \is gate represented by the point 
\(B=[\pi/4, \pi/4, 0]\) in Fig.~\ref{2D}:
\begin{align}
\sqrt{iSWAP} =\left[\begin{array}{rrrr}
1&0&0&0\\
0&\frac{1}{\sqrt{2}}&\frac{i}{\sqrt{2}}&0\\
0&\frac{i}{\sqrt{2}}&\frac{1}{\sqrt{2}}&0\\
0&0&0&1
\end{array}\right],
\end{align}
and \rs gate represented by the point $B_3=[\pi/4, \pi/4, \pi/4]$ in 
Fig.~\ref{Weyl-Chamber}: 
\begin{align}
    &\sqrt{SWAP} =\left[\begin{array}{rrrr}
1&0&0&0\\
0&\frac{1+i}{2}&\frac{1-i}{2}&0\\
0&\frac{1-i}{2}&\frac{1+i}{2}&0\\
0&0&0&1
\end{array}\right].
\end{align}
Each of these gates together with some single-qubit gates can 
construct a universal gate set.

Recall that we cannot determine the Cartan decomposition \eqref{cartan} 
uniquely, for any two-qubit unitary matrix $U$. 
Thus, we used the decomposition algorithm 'TwoQubitBasisDecomposer' 
implemented in Qiskit \cite{Qiskit}. 
Figure~\ref{SQISWAP_CIRC} shows two types of decomposed gate layout of 
\is based on CX (middle) and CV (lower), which we call \isx and \isv, 
respectively. 
Also the case of \rs is shown in Fig.~\ref{SQSWAP_CIRC}, where the 
CX- and CV-based decompositions are called \rsx and \rsv, respectively.
Here, $U_2(\phi,\lambda)$ and $U_3(\theta, \phi, \lambda)$ are the single qubit gates in the QASM language~\cite{cross2017open}, defined as follows:
\begin{align}
U_2(\phi,\lambda)=
\frac{1}{\sqrt{2}}
\begin{bmatrix}
1 & -e^{i\lambda} \\
e^{i\phi} & e^{i(\phi + \lambda)} \\
\end{bmatrix},\\
U_3(\theta, \phi, \lambda)=
\begin{bmatrix}
\cos(\theta/2) & -e^{i\lambda}\sin(\theta/2) \\
e^{i\phi}\sin(\theta/2) & e^{i(\phi + \lambda)}\cos(\theta/2) \\
\end{bmatrix}
\end{align}

The pulse schedules corresponding to these four decomposed circuits are 
shown in Figs.~\ref{sqi_pulse} and \ref{sq_pulse}, where the pulse for 
CX and CV were implemented with the optimized CR duration time identified 
in Section~\ref{sec_cv}.

\begin{figure}[ht]
\vspace{-10pt}
    \begin{minipage}{\hsize}
    \centering
    \scriptsize
\[\Qcircuit @C=1em @R=1.2em {
& \multigate{1}{\sqrt{iSWAP}} & \qw \\
& \ghost{\sqrt{iSWAP}}& \qw
}\]
\vspace{2pt}
    \end{minipage}
    \begin{minipage}{\hsize}
    \centerline{\rotatebox[origin=c]{90}{$\equiv$}}
    \scriptsize
\[\Qcircuit @C=1em @R=1.2em {
& \gate{U_2(\frac{-3\pi}{2}, \frac{\pi}{2})} &  \ctrl{1} & \gate{U_3(\frac{3\pi}{4}, 0,\frac{3\pi}{2})} & \ctrl{1} & \gate{U_2(\frac{3\pi}{2},0)}& \qw\\
& \gate{U_2(\frac{-\pi}{2},\frac{\pi}{2})}& \targ & \gate{U_2(\frac{3\pi}{2},\frac{\pi}{4})} & \targ & \qw & \qw
}\]
\vspace{2pt}
    \end{minipage}
    \begin{minipage}{\hsize}
    \centerline{\rotatebox[origin=c]{90}{$\equiv$}}
    \scriptsize
\[\Qcircuit @C=1em @R=1.2em {
& \gate{U_2(0,0)} &  \ctrl{1} & \gate{U_2(\frac{\pi}{2}, \frac{5\pi}{4})} & \ctrl{1} & \gate{U_2(\frac{3\pi}{2}, \frac{3\pi}{4})}& \qw\\
& \gate{Z}& \gate{V} & \gate{U_3(\frac{\pi}{4}, 0,\frac{-\pi}{2})} & \gate{V} & \gate{U_3(\frac{\pi}{4}, 0,\frac{-\pi}{2})} & \qw
}\]
    \end{minipage}
    \caption{Circuit diagram for \isx gate~(middle) and \isv gate~(lower).}
    \label{SQISWAP_CIRC}
\end{figure}

\begin{figure}[ht]
\vspace{-10pt}
    \begin{minipage}{\hsize}
    \centering
    \scriptsize
\[\Qcircuit @C=1em @R=1.2em {
& \multigate{1}{\sqrt{SWAP}} & \qw \\
& \ghost{\sqrt{SWAP}}& \qw
}\]
\vspace{2pt}
    \end{minipage}
    \begin{minipage}{\hsize}
    \centerline{\rotatebox[origin=c]{90}{$\equiv$}}
    \tiny
\[\Qcircuit @C=1em @R=1.2em {
& \gate{Z} & \ctrl{1} & \gate{U_3(\frac{\pi}{4},\frac{-3\pi}{2},\frac{\pi}{2})} &  \ctrl{1} & \gate{U_3(\frac{\pi}{4},0,\frac{-3\pi}{2})} & \ctrl{1} & \gate{Z}& \qw\\
& \gate{U_3(\pi, \pi, -\pi)} & \targ & \gate{U_2(-\pi, \frac{-3\pi}{4})}& \targ & \gate{U_2(0, \frac{-3\pi}{2})} & \targ & \gate{U_2(\pi, \frac{\pi}{2})} & \qw
}\]
\vspace{2pt}
    \end{minipage}
    \begin{minipage}{\hsize}
    \centerline{\rotatebox[origin=c]{90}{$\equiv$}}
    \tiny
\[\Qcircuit @C=1em @R=1.2em {& \gate{U_2(\pi,0)}& \ctrl{1} & \gate{U_2(\frac{\pi}{2}, \frac{5\pi}{4})} &  \ctrl{1} & \gate{U_2(\frac{-\pi}{2}, \frac{3\pi}{4})}  & \ctrl{1} & \gate{U_3(0,\frac{\pi}{2},\frac{-5\pi}{4})}& \qw\\& \gate{X} & \gate{V} & \gate{U_3(\frac{\pi}{4},0,\frac{-\pi}{2})}& \gate{V} & \gate{U_3(\frac{3\pi}{4},\pi,\frac{-\pi}{2})}& \gate{V} & \gate{U_2(\frac{7\pi}{4}, \pi)} & \qw}\]
    \end{minipage}
    \caption{Circuit diagram for \rsx gate~(middle) and \rsv gate~(lower).}
    \label{SQSWAP_CIRC}
\end{figure}

\label{app_swappulse}
\begin{figure}[ht]
    \begin{minipage}{\hsize}
    \centering
    \includegraphics[bb=80 0 1100 580.0,clip,width = 195pt]{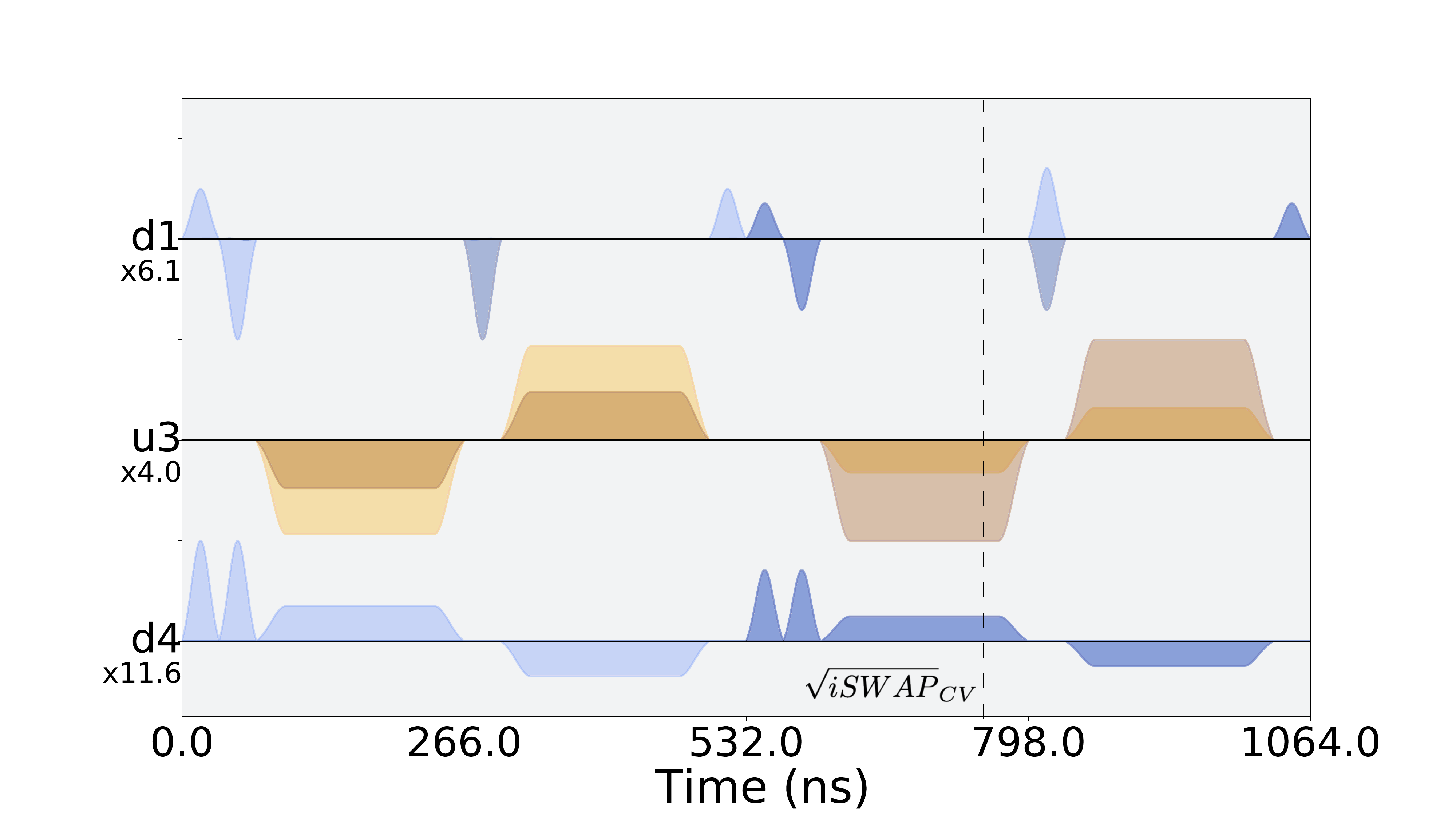}
    \end{minipage}
    \begin{minipage}{\hsize}
    \centering
    \includegraphics[bb=80 0 1100 580.0,clip,width = 195pt]{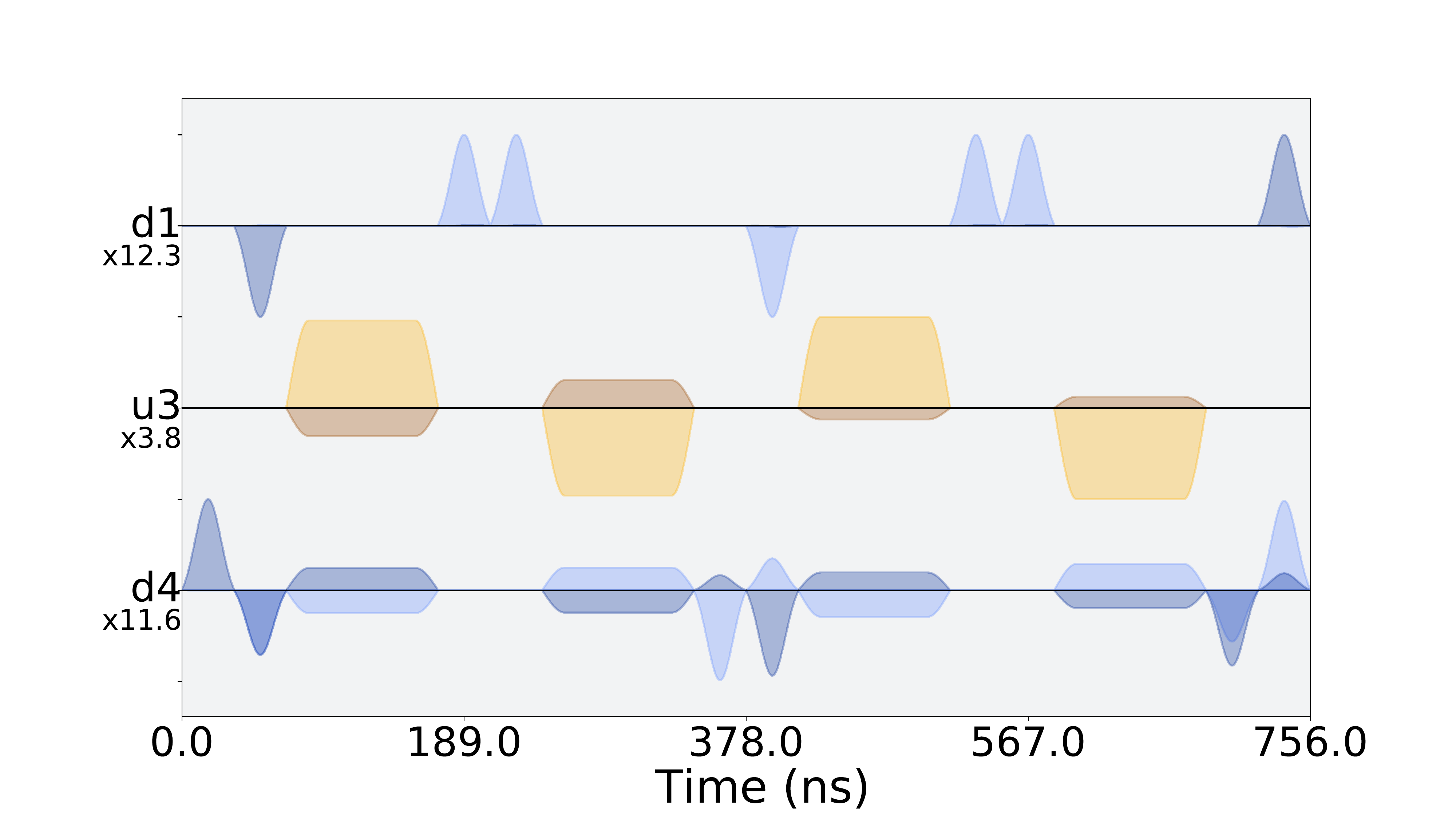}
    \end{minipage}
    \caption{Pulse sequences of \isx (upper) and \isv (lower), corresponding 
    to the circuit diagram in Fig.~\ref{SQISWAP_CIRC}.Dashed line denotes the total gate time of \isv.}
    \label{sqi_pulse}
\end{figure}

\begin{figure}[ht]
    \begin{minipage}{\hsize}
    \centering
    \includegraphics[clip,width = 220pt]{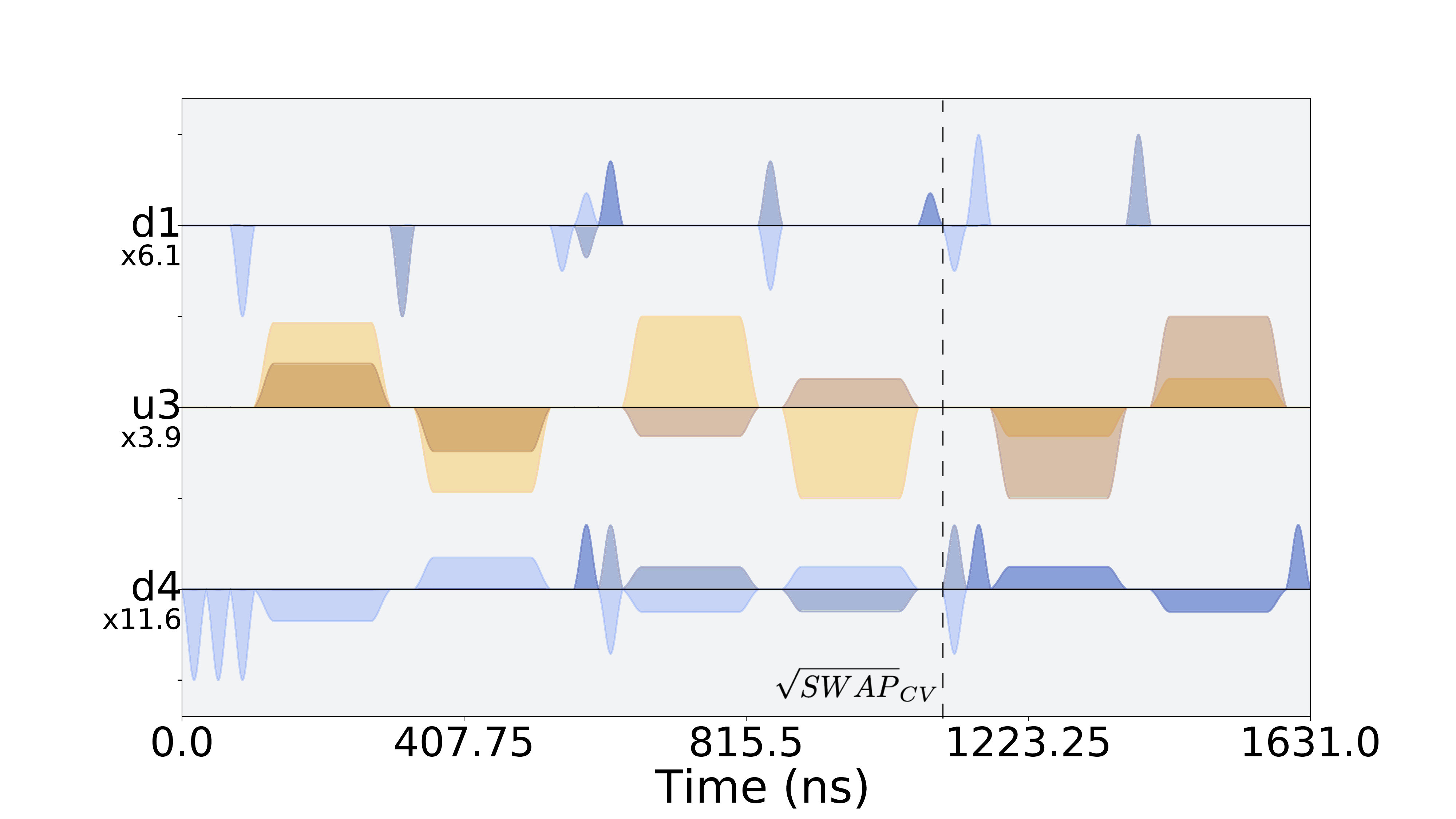}
    \end{minipage}
    \begin{minipage}{\hsize}
    \centering
    \includegraphics[clip,width =220pt]{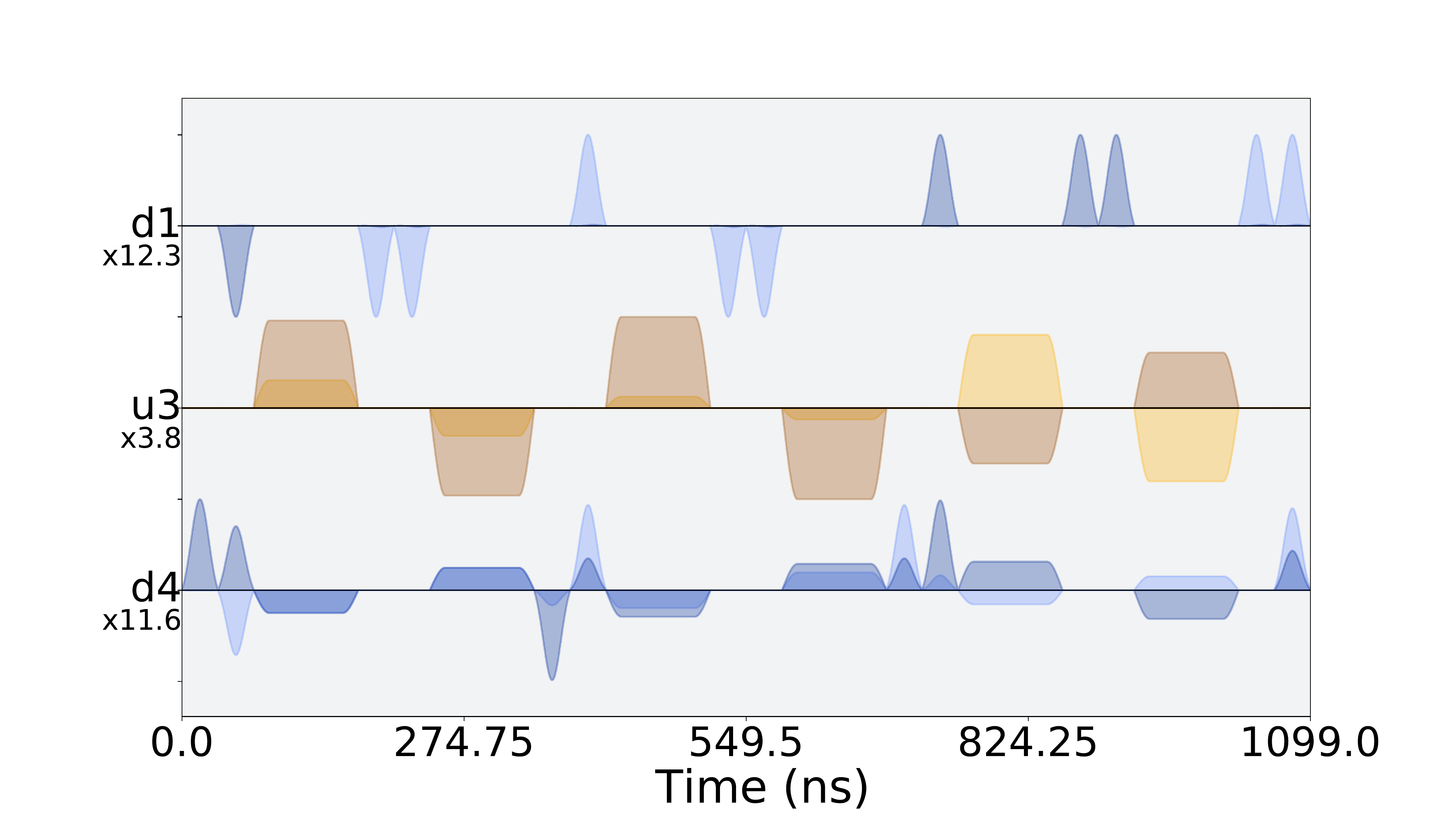}
    \end{minipage}
    \caption{Pulse sequences of \rsx (upper) and \rsv (lower), corresponding 
    to the circuit diagram in Fig.~\ref{SQSWAP_CIRC}. }
    \label{sq_pulse}
\end{figure}

Figure~\ref{sqi_pulse} shows that the total gate time of \isv is 
shortened by 308 ns compared to that of \isx. 
Also, from Fig.~\ref{sq_pulse} we find that the total gate time of 
\rsv is $532$ ns shorter than that of \rsx. 
We compute the gate fidelities of these four gates to their ideal 
correspondence, using QPT. 
The results are summarized in Table~\ref{swaptable}, together with 
the gate time; 
\is (\rs) gate with CV gates achieves the better fidelity by 0.87 
(2.14)~\% compared with the default CX-based implementation. 
This might be thanks to the shortened gate time realized via the 
pulse-engineered CV gate. 
Note that, when \is or \rs is involved in some larger quantum circuits, 
the gate-time advantage of the CV-based implementation may lead to 
significant improvement in the fidelity of those circuit.

\begin{table}[ht]
 \centering
 \caption{
 Gate fidelity and gate time of $\sqrt{iSWAP}$ and $\sqrt{SWAP}$, 
 implemented with the default CX gate or the pulse-engineered CV gate. 
 For each input, we performed 8192 shots and calculated the gate fidelity 
 $F_P$ with the use of read-out error mitigation.}
 \centering
 \begin{tabular}{ccccc} 
    Gate &  \#CX & \#CV & Fidelity ($F_{P}$) & Gate time (ns)\\\hline
  $\isx$ & 2 & -- & 0.9765 & 1064\\ 
  $\isv$ & -- & 2 & 0.9852 & 756\\ 
  \rsx & 3 & -- & 0.9604 & 1631\\ 
  \rsv & -- & 3 & 0.9818 & 1099\\ 
 \end{tabular}
  \label{swaptable}
\end{table}

\section{High-speed and high-precision Toffoli gate with CV gates}

The pulse-engineered CV gate can be applied to improve the speed and 
precision of bigger size gates beyond the two-qubit case. 
As a demonstration, here we study the three-qubit Toffoli gate (or 
the Controlled-Controlled-X gate). 
The idea presented here is applicable to the general multi-qubit 
Toffoli gate appearing in many long-term algorithms such as QRAM 
database~\cite{giovannetti2008quantum} and the diffusion operator 
in Grover's search algorithm~\cite{grover1997quantum}. 

\subsection{Gate implementation for linearly-coupled three qubits}
\begin{figure}[ht]
    \begin{minipage}{\hsize}
    \[\Qcircuit @C=.5em @R=.5em @!R {
\lstick{q_0}&\qw & \ctrl{1} & \qw & \targ & \ctrl{1} &\qw&\qw&\rstick{q_1}\\
\lstick{q_1}&\ctrl{1} & \targ & \ctrl{1} & \ctrl{-1} &\targ &\ctrl{1}&\qw&\rstick{q_0}\\
\lstick{q_4}&\gate{V} & \qw &
\gate{V^{\dag}} & \qw &\qw &\gate{V} & \qw&\rstick{q_4}
}\]
    \end{minipage}
    \begin{minipage}{\hsize}
    \small
    \[\Qcircuit @C=.25em @R=.5em @!R {
\lstick{q_0}&\qw&\gate{T^\dagger} & \ctrl{1} &  \qw & \ctrl{1} & \qw & \ctrl{1} & \qw &\qw& \qw & \ctrl{1} & \qw & \qw&\qw&\qw&\rstick{q_0}\\
\lstick{q_1}&\qw& \gate{T^\dagger} & \targ & \ctrl{1} & \targ &\ctrl{1} & \targ &\gate{Z} &\gate{T} &\ctrl{1}& \targ & \qw& \ctrl{1}& \qw&\qw&\rstick{q_1}\\
\lstick{q_4}&\gate{H}&\gate{T^\dagger}& \qw &\targ &  \gate{T^\dagger} &\targ & \gate{Z} &\gate{T} &\qw&\targ & \gate{Z} & \gate{T}&\targ&\gate{H}&\qw&\rstick{q_4}}\]
    \end{minipage}
    \caption{Toffoli gate for linearly coupled three qubits. 
    (Upper) \tv composed of CV and CX gates. 
    (Lower) \tx composed of only CX gates \cite{gwinner2020benchmarking}. 
    Note that the former exchanges $q_0$ and $q_1$, while maintaining 
    the functionality of Toffoli gate.}
    \label{toffoli_circ}
\end{figure}
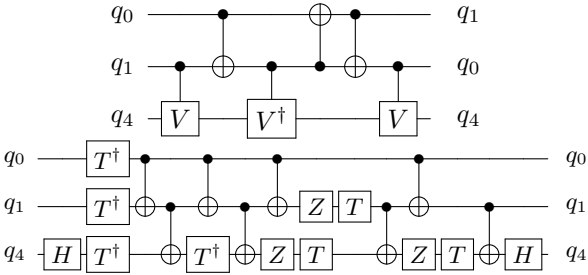

If three qubits are fully connected, then we can construct Toffoli gate 
using 6 CX gates (and some single-qubit gates), while the combination 
of 3 CV and 2 CX gates also constructs Toffoli gate; hence the 
pulse-engineered CV gate enables reducing the total gate time. 
However, the standard structure of the current IBM Quantum devices is 
of the linear coupling form of qubits, in which case the number of necessary gates increase.

Here we consider two different construction of Toffoli gate with and 
without CV gates, \tv and \tx gate shown in Fig.~\ref{toffoli_circ}; 
note that $q_j~(j=0,1,4)$ represents the $j$th qubit of ibmq\_tronto 
device shown in Fig.~\ref{processor} and thus $q_0$ and $q_4$ are not 
directly connected. 
\tv gate has 3 CX and 3 CV gates; hence, with the use of 
pulse-engineered CV gate, the total gate time of \tv becomes shorter 
than that of the textbook Toffoli with 6 CX gates as well as \tx. 
Note that the SWAP gate is built in there to connect $q_0$ and $q_4$; 
consequently, \tv exchanges $q_0$ and $q_1$, while maintaining the 
functionality of Toffoli gate. 
However, the pure Toffoli composed of \tv and subsequent SWAP gate 
needs 6 CX and 3 CV gates, meaning that it still can be realized with 
shorter gate time than \tx by the pulse engineering of CV.

\subsection{Experimental results}

\begin{figure}[ht]
    \begin{minipage}{\hsize}
    \centering
    \includegraphics[bb=80 0 1100 580.0,clip,width = 240pt]{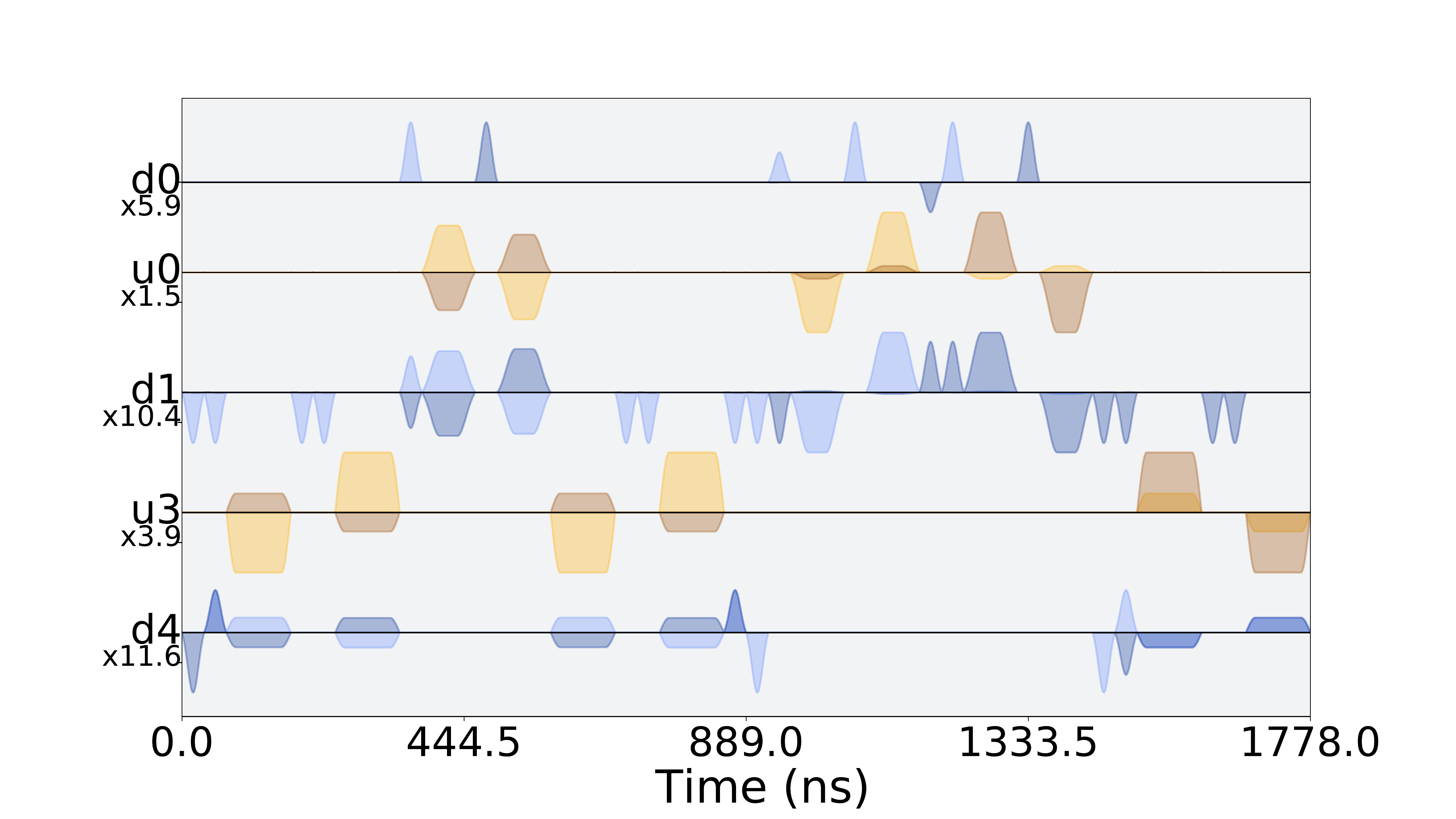}
    \end{minipage}
    \begin{minipage}{\hsize}
    \centering
    \includegraphics[bb=80 0 1100 580.0,clip,width = 240pt]{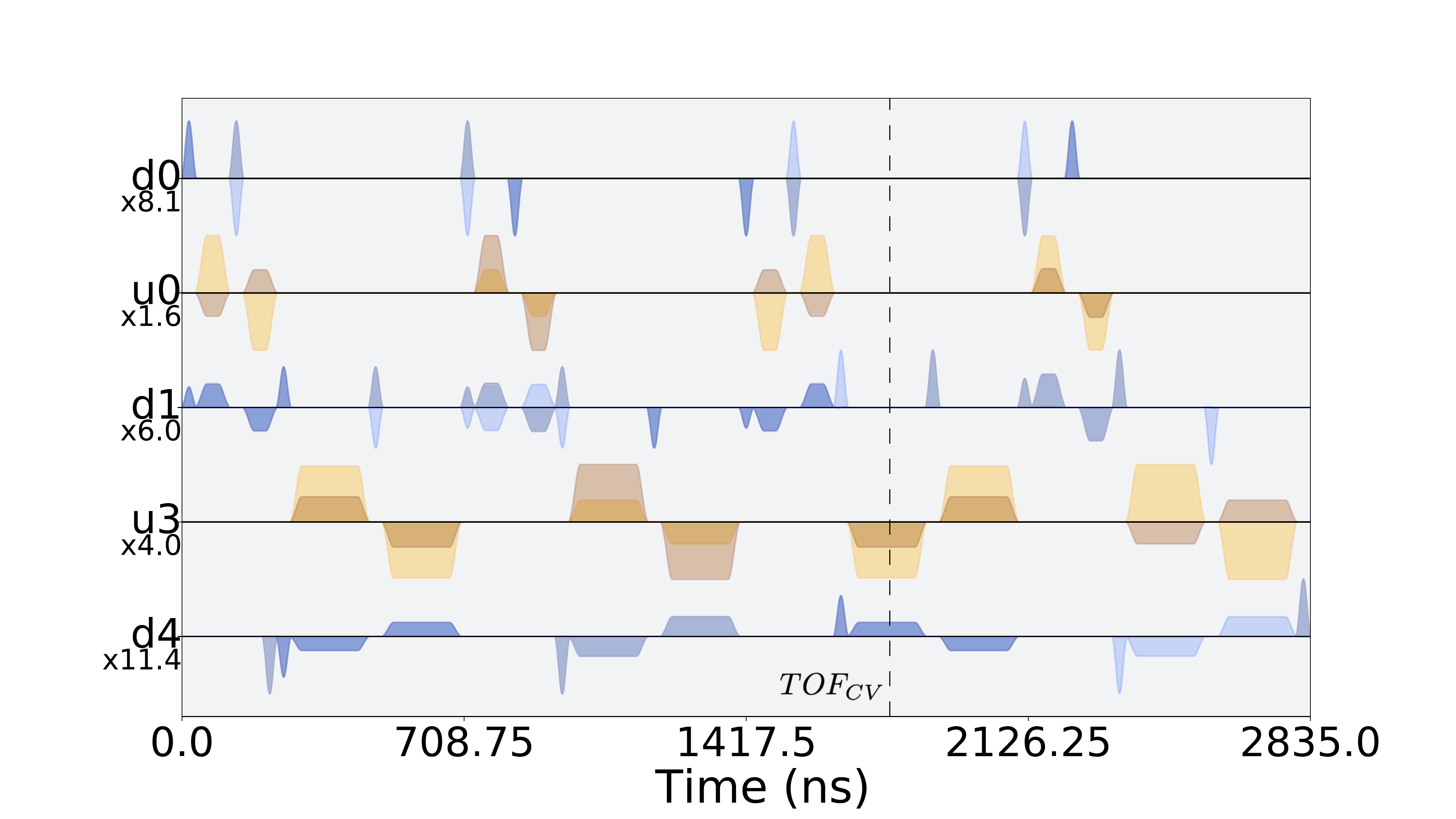}
    \end{minipage}
    \caption{Pulse sequences of \tv composed of CX and CV gates (upper) 
    and that of \tx composed of only CX gates (lower). 
    In addition to the channels used in the previous sections, DriveChannel d0 of the control qubit and ControlChannel u0 as interaction of qubit 0 and qubit 1  are used. 
    The gate time of \tv and \tx are 1778 ns and 2835 ns, respectively. }
    \label{toffoli_pulse}
\end{figure}

We conducted an experiment to compare the actual performance of 
\tv (3 CX and 3 CV) to \tx (8 CX), where the pulse-engineered CV 
is used in the former, on ibmq\_tronto processor shown in 
Fig.~\ref{processor}. 
The pulse sequences corresponding to these Toffoli gates are 
depicted in Fig.~\ref{toffoli_pulse}. 
As expected, the total gate time are 1778 ns and 2835 ns for \tv 
and \tx, respectively, suggesting that \tv would have better 
precision than \tx.

Since QPT requires an excessive number of experiments, we have 
adopted the quantum state tomography and calculated the state fidelity 
\cite{magesan2011gate}:
\begin{equation}
   F_{s}(\rho_{\rm exp}, \rho_{\rm ide}) = 
       Tr[\sqrt{\sqrt{\rho_{\rm exp}}\rho_{\rm ide}\sqrt{\rho_{\rm exp}}}]^2,
\label{fidelitysiki2}
\end{equation}
where $\rho_{\rm ide}$ denotes the ideal target density matrix and 
$\rho_{\rm exp}$ denotes the reconstructed density matrix in the 
experiment using the state tomography. 
As the input state to Toffoli gate, we prepared 12 states listed in 
Table~\ref{Toffolitable}, where $\ket{\pm}=(\ket{0}\pm\ket{1})/\sqrt{2}$. 
We performed 8192 shots (measurements) for each initial state and 
calculated $F_{s}(\rho_{\rm exp}, \rho_{\rm ide})$. 
Table~\ref{Toffolitable} summarizes the results, showing the superiority 
of \tv for all input states except $\ket{111}$. 
As a result, $TOF_{CV}$ has 4.06~\% higher average fidelity than \tx. 
This is a bigger superiority of the CV-based gate over the conventional 
CX-based one, compared to the previous case shown in Table~1, simply 
because the gate length becomes longer.

\begin{table}[ht]
 \centering
 \caption{
 Fidelity $F_{\text{s}}$ for the two types of Toffoli gates. }
  \begin{tabular}{c|cc} 
  \label{Toffolitable}
    Input states & \tx  & \tv  \\\hline
    $\ket{000}$ & 0.9287 & 0.9773\\ \hline
    $\ket{001}$ & 0.9538 & 0.9711\\ \hline
    $\ket{010}$ & 0.8886 & 0.9318\\ \hline
    $\ket{011}$ & 0.9344 & 0.9450\\ \hline
    $\ket{100}$ & 0.8697 & 0.9259\\ \hline
    $\ket{101}$ & 0.9113 & 0.9481\\ \hline
    $\ket{110}$ & 0.9167 & 0.9284\\ \hline
    $\ket{111}$ & 0.9217 & 0.9089\\ \hline
    $\ket{+10}$ & 0.8454 & 0.9462\\ \hline
    $\ket{1+0}$ & 0.9046 & 0.9341\\ \hline
    $\ket{++1}$ & 0.8860 & 0.9513\\ \hline
    $\ket{--1}$ & 0.8603 & 0.9408\\ \hline
    Average & 0.9018 & 0.9424\\ 
  \end{tabular}
\end{table}

\section{Conclusion}

Using only CX gates for entangling qubits in quantum computation is now 
a \textit{de facto} standard. 
IBM Quantum is no exception. 
While this approach is less burdensome for calibration, it has the disadvantage 
that some gate/circuit structure become redundant. 
To resolve this issue, in this paper we proposed using CV gates in addition to 
the default gate set; actually OpenPulse allows us to realize CV gate with 
shorter gate time than that of CX gate as well as the default CV gate composed 
of 2 CX gates. 
The parameters of the corresponding CR Hamiltonian for realizing such 
pulse-engineered CV gate are the same as those of the CX gate, except for 
the pulse length and some local gate parameters, meaning that the calibration 
burden is not significant. 
In particular, the result of Section II (Fig.~4) indicates that the optimal 
pulse length does not change in each calibration. 
The gate-time improvement in circuit design, which eventually leads to the 
gate-fidelity improvement, has been demonstrated with \(\sqrt{SWAP}\), 
\(\sqrt{iSWAP}\), and Toffoli gates. 
Note that the gate fidelity improvement were not totally great (0.66\% improvement 
for the CV implementation, 0.87\% for $\sqrt{iSWAP}$, and 2.14\% for $\sqrt{SWAP}$), 
and this may be due to the presence of ZZ interactions that cannot be counteracted 
by the echo scheme~\cite{jikken} that was employed in our method. 
Suppression of the ZZ interactions~\cite{kandala2011demonstration,PRXQuantum.1.020318,mitchell2021hardware,wei2021quantum} would allow us to further improve the gate performance.

In summary, from the practicality and feasibility viewpoint, we believe that 
the new gate set that contains the proposed pulse-engineered CV gate can be 
used to effectively reduce the redundancy of several quantum circuits, thereby 
realize shorter gate time in total, and eventually improve several quantum 
algorithms. 
Actually, to investigate a wider range of applications, we plan to 
execute comparative verification of the proposed method on a bigger-size 
circuit or a near-term quantum algorithm.

\appendix
\section{}
\subsection{Gaussian square pulse envelope}
\label{sec_envelope}

In all experiments we employed the Gaussian-Square pulse composed 
of the constant-amplitude part of length (width) \(\tau_w\) and 
Gaussian-formed rising and falling edges of length \(\tau_r\). 
The overall pulse waveform $f(t)$, as a function of time $t$, is 
thus given by 
\begin{align*}
    f(t) = 
    \begin{cases}
       A\exp( -\frac{1}{2\sigma^2}(t - \frac{\tau_r}{2})^2), ~~
                (0 \leq t < \tau_r) \\
       A, ~~ (\tau_r \leq t < \tau_r + \tau_w) \\
       A\exp( -\frac{1}{2\sigma^2}(t -\frac{(\tau_r+\tau_w)}{2})^2), ~~ 
                (\tau_r + \tau_w \leq t < \tau_d), \\
  \end{cases}
\end{align*}
where $A$ is the maximum amplitude and $\sigma^2$ is the variance 
of the Gaussian part, respectively. 
Note that the overall pulse length or the duration is defined as 
\begin{align}
    \tau_d = 2\tau_r+\tau_w. 
\label{length}
\end{align}
%

\subsection{Optimal pulse duration of CX gate}
\label{sec_cxpulse}

Figure 1 shows the pulse schedule for implementing CX gate, where the CR 
pulse duration is 196 ns and accordingly the total gate time 462 ns; 
this is actually the best value that achieves the maximum gate fidelity. 
Here we show the detail of the OpenPulse experiment to identify this 
optimal duration. 

The experiment was conducted in the same setting described in 
Section II-C, with the use of qubit 1 and 4. 
We evaluated the following trial CX gate with changing the duration 
$\tau_d \in [144, 259]$ ns: 
\begin{equation}
    CX_{\rm{trial}}(\tau_d) = [ZI]^{1/2}[ZX]^{\theta(\tau_d)}[IX]^{1/2}, 
\end{equation}
where $\theta(\tau_d)=-\tau_d / 2\tau_{CX}$ with the nominal value 
$\tau_{CX}=196$ ns. 
The yellow and blue lines in Fig.~\ref{cx_fidelity} depict the gate 
fidelity between the pulse-engineered $CX_{\rm{trial}}(\tau_d)$ and 
the ideal CX gate, with and without the readout error mitigation 
respectively. 
The black dotted line depicts the gate fidelity between the 
theoretical $CX_{\rm{trial}}(\tau_d)$ and the ideal CX gate. 
The figure thus shows that the optimal duration is exactly the 
nominal value, i.e., $\tau_d=\tau_{CX}=196$ ns, which achieves 
the perfect gate fidelity.

\Figure[htb][width = 240pt]
{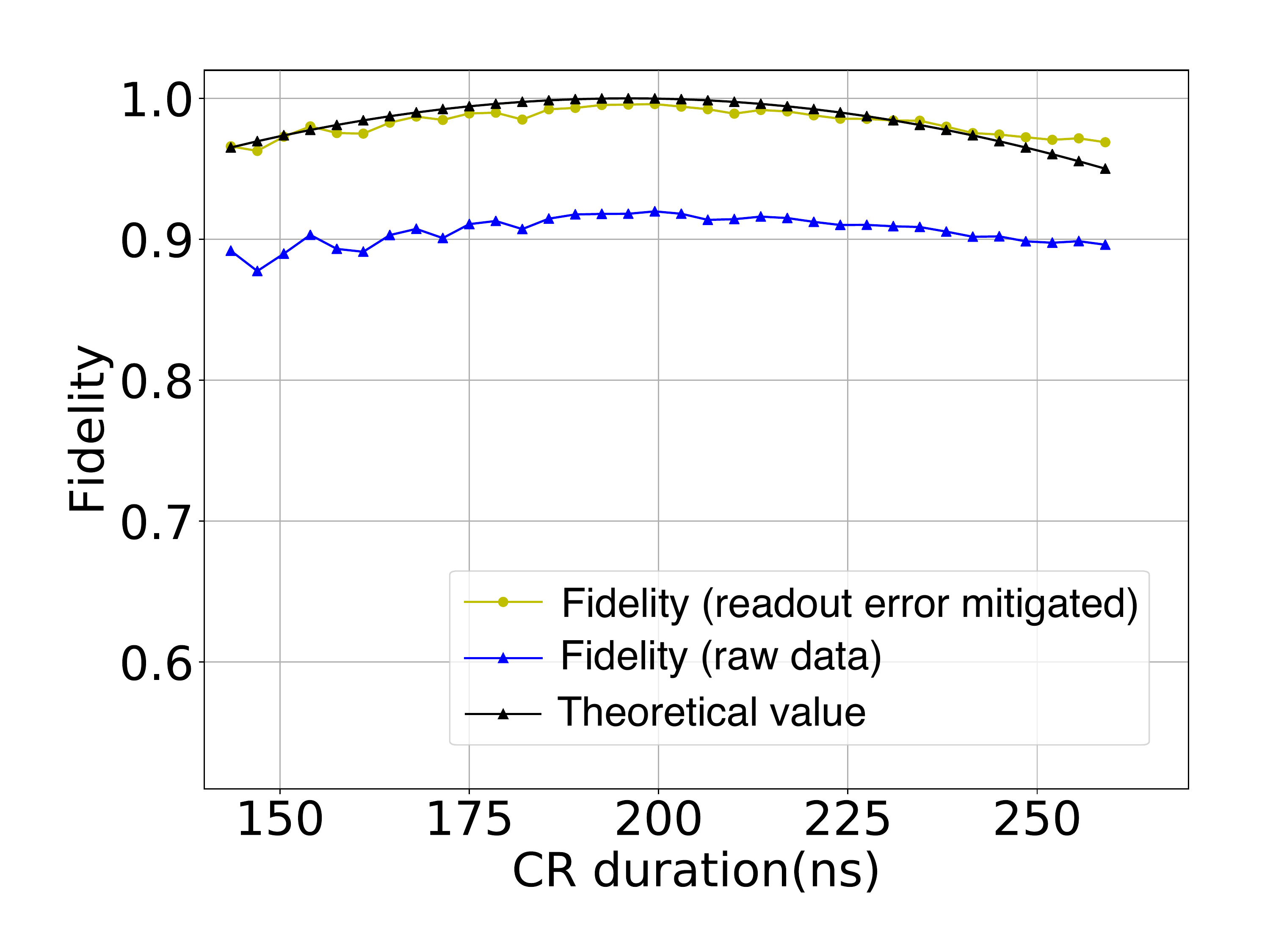}
{Gate fidelity of the trial CX gate (17) to the ideal CX gate, 
as a function of the duration of CR pulse. 
\label{cx_fidelity}}

\section*{Acknowledgment}
The results presented in this paper were obtained in part using 
an IBM Q quantum computing system as part of the IBM Q Network. 
The views expressed are those of the authors and do not reflect the
official policy or position of IBM or the IBM Q team. 

\bibliographystyle{IEEEtran}
\bibliography{openpulse}

\EOD

\end{document}